\documentclass[[final,3p,times,twocolumn]{elsarticle}

\usepackage{graphicx}% Include figure files
\usepackage{dcolumn}% Align table columns on decimal point
\usepackage{bm}% bold math
\usepackage{amsmath}
\usepackage{comment}
\usepackage{epstopdf} %converting to PDF
\usepackage{lineno,hyperref}
\usepackage{color} %for the comments by Jo
\usepackage{soul} %Abner's comments in highlighted text
\usepackage{ulem} %(Abner) to strikethrough text

\newcommand{\chem}[1]
{
\ensuremath{\mathrm{#1}}
}

\usepackage{amssymb}

\usepackage{lineno,hyperref}
\modulolinenumbers[5]

\journal{Ultramicroscopy}

\begin{document}

\begin{frontmatter}



\title{Reducing electron beam damage through alternative STEM scanning strategies. Part I -- Experimental findings}

\author[emat,nanolab]{A. Velazco}
\author[emat,nanolab]{A. B\'ech\'e}
\author[emat,nanolab]{D. Jannis}
\author[emat,nanolab]{J. Verbeeck\corref{mycorrespondingauthor}}
\cortext[mycorrespondingauthor]{Corresponding author}
\ead{jo.verbeeck@uantwerp.be}

\address[emat]{EMAT, University of Antwerp, Groenenborgerlaan 171, 2020 Antwerp, Belgium}
\address[nanolab]{NANOlab Center of Excellence, University of Antwerp, Groenenborgerlaan 171, 2020 Antwerp, Belgium}

\begin{abstract}
The highly energetic electrons in a transmission electron microscope (TEM) can alter or even completely destroy the structure of samples before sufficient information can be obtained. This is especially problematic in the case of zeolites, organic and biological materials. As this effect depends on both the electron beam and the sample and can involve multiple damage pathways, its study remained difficult and is plagued with irreproducibity issues, circumstantial evidence, rumours, and a general lack of solid data. 
Here we take on the experimental challenge to investigate the role of the STEM scan pattern on the damage behaviour of a commercially available zeolite sample with the clear aim to make our observations as reproducible as possible. We make use of a freely programmable scan engine that gives full control over the tempospatial distribution of the electron probe on the sample and we use its flexibility to obtain mutliple repeated experiments under identical conditions comparing the difference in beam damage between a conventional raster scan pattern and a newly proposed interleaved scan pattern that provides exactly the same dose and dose rate and visits exactly the same scan points. We observe a significant difference in beam damage for both patterns with up to 11 \% reduction in damage (measured from mass loss). These observations demonstrate without doubt that electron dose, dose rate and acceleration voltage are not the only parameters affecting beam damage in (S)TEM experiments and invite the community to rethink beam damage as an unavoidable consequence of applied electron dose.
\end{abstract}


%\begin{itemize}
%\item An alternative interleaving pattern is compared to the conventional raster pattern in terms of beam damage. Both methods provide the same electron dose and spatial resolution. However, a different distribution of the dose in time and space is achieved with the alternative pattern. 
%\item Automated experiments on a commercial zeolite sample were performed to increase the reproducibility of the results.
%\item The methodical experiments show that the alternative scanning, compared to the raster scanning, systematically reduces electron beam damage.  
%\item A general diffusion model is employed in an attempt to empirically model the observed damaging process.
%\end{itemize}


\end{frontmatter}


\section{Introduction}
For many important classes of materials, e.g. zeolites, organic and biological materials, electron beam damage in transmission electron microscopy (TEM) is a detrimental effect that limits the capabilities to obtain atomic scale information. Displacement of atoms and consequent structure degradation is the result of the interaction between the highly energetic electrons and beam sensitive materials in TEM. These classes of materials, along with many others classified as insulators, predominantly suffer from radiolysis or ionization damage, unlike conductive materials that are mainly affected by knock-on damage \cite{Egerton2004,Egerton2013}. Independently of the operation mode of the microscope (scanning or transmission mode), commonly, ionization damage (hereafter beam damage) is considered to be proportional to the electron dose that is delivered to the sample. Hence, conventional strategies to reduce beam damage in TEM consist in reducing the electron dose either by lowering the electron beam current or by making short exposure time images. Both generally result in low signal-to-noise (SNR) ratio images and further restrict the accuracy and precision of the quantified parameters extracted from such data.\\
However, experimental indications suggest that the total electron dose is not the only factor to consider. In scanning transmission electron microscopy (STEM), dose rate (electron dose per unit of time) thresholds and dose fractionation have been found to be ways of overcoming or reducing the accumulation of damage. A.C. Johnston-Peck et al. \cite{Johnston-Peck2016} reported a dose rate threshold for cerium dioxide for which no structural changes were observed in the STEM images as well as in the \chem{Ce} \chem{M_{4,5}} edges monitored by electron energy loss spectroscopy (EELS). L. Jones et al. \cite{Jones2018} compared different acquisition methods on a lead perovskite (\chem{Pb_2ScTaO_6}) under a fixed total dose and dose rate. Fast scanning multi-frame acquisitions with equal dose per frame provided reduced sample degradation compared to a single acquisition in STEM imaging and Spectrum Imaging (SI). \\
Moreover, it is well known that beam damage can extend over an area larger than the size of the irradiated area \cite{Jiang2002,Jiang2017}. This effect may have many causes which can itself depend heavily on the material: delocalised inelastic scattering, fast secondary electrons, generation and diffusion of radicals, diffusion of heat, electrostatic charge and dielectric breakdown, etc. \cite{Jiang2017, Lutwyche1992, Egerton2012, Wu2001}. \\
Most of these effects are triggered primarily by ionization where a fraction of the energy from the incoming electrons is transferred to the sample with a subsequent energy transformation which can be recurrent during damaging until the energy is dissipated. This mechanism is dynamic in nature and has both a spatial and temporal scaling parameter that we will attempt in the second part of this manuscript to describe as a diffusion process as e.g. also attempted by D. Nicholls et al. \cite{Nicholls2020}.\\
 In STEM, where a highly focused electron probe interacts with the sample, the conventional ‘raster’ scanning applies the electron dose in a line by line fashion where adjacent pixels are scanned consecutively in each line. A damage process that spreads spatially with time, can affect regions of the sample that will be visited by the next probe position. In this way the damage could build up rapidly as regions that come later in the scan will have been affected more by earlier scan points. For similar reasons, scan positions with more neighbours (central region of a scanned area) will suffer more than positions which have less neighbours (edges of the scanned area).\\
A more recent non-conventional strategy to reduce beam damage in TEM involves a compressed sensing (CS) approach. In the theory of CS a faithful representation of a signal can be retrieved from a random undersampling data acquisition under the assumption that the signal has a sparse representation in a properly chosen basis \cite{Donoho2006}. Applied to STEM, it is possible to obtain a faithful restoration of an image by scanning only a subset of the scan positions of a conventional acquisition, and atomic-resolution imaging with as little as 20\% of the pixels was demonstrated \cite{Beche2016,Kovarik2016}. Although from a statistical point of view it was found that for Poisson noise limited imaging, CS does not present any improvement when compared to standard sampling with equivalent electron dose \cite{VandenBroek2019}. Experimentally, we and others observe however a qualitative improvement in beam damage which seems to contradict this theoretical finding. A possible explanation for this observation is that the way the electron dose is spatially and temporally distributed on the sample does matter for beam damage. Indeed, in CS acquisition, the average distance between consecutive pixels is larger than in conventional scanning. Intuitively, one could imagine that this larger distance between sampled points could prevent the accumulation of damage by outrunning the diffusion-like effects coming from earlier scanning positions as hypotesised above.\\
Similar concepts were explored in the field of scanning electron microscopy (SEM) and low-voltage electron beam lithography (LVEBL) to reduce the undesirable effects of electrostatic charging in insulators \cite{Thong2006,Mun2004}. Changing the sequence of the scanning positions in SEM and the sequence of the written patterns in LVEBL allowed to counteract the charging effects that otherwise build up when immediately adjacent positions/patterns are scanned/written; taking advantage of the characteristic time decay of electrostatic charging depending on conductivity.\\
In STEM hyperspectral imaging, A. Zobelli et al. \cite{Zobelli2020} have recently investigated a random scan operation mode in order to reduce beam damage  effects. The effect of the scanning pattern was shown in the cathodoluminescence map of hexagonal boron nitride flakes which exhibited intensity instabilities. The instabilities were suggested to come from variations in the charge state of defect centers, which can be influenced by the accumulation of electrostatic charge in the sample caused by the ejection of secondary electrons. In this sense, raster scanning would generate a fast accumulation of electrostatic charge as adjacent pixels are scanned consecutively. Intensity instabilities were reduced when scanning with a random pattern. \\
Allowing for the possibility that beam damage can be related to and mediated by a diffusion process triggers the question whether different scanning strategies could offer reduced beam damage while keeping the same image quality and total dose on the sample. \\
In this first part of the manuscript, we will experimentally investigate the role of an alternative interleaved scan pattern on beam damage. We propose an alternative interleaved scan pattern that is sketched in Figure \ref{scansketch}. The scanning is done without any interruption which allows achieving the same acquisition time (and dose) than for the conventional raster method. The maximum number of pixels that are skipped in each scanning direction is limited by the dynamic performance of the scan system where magnification and dwell time play a decisive role as they determine the settling time of the probe to within an acceptable region from the new probe position. More importantly, the number of pixels to skip would be dictated by the diffusion parameter that governs the beam damage process. We attempt to make the distance between consecutively visited sample positions such as to ensure that a newly visited position is not yet influenced by the diffusing effect from the previous one. This makes finding an optimal alternative scanning pattern a non-trivial problem and likely dependent on the sample characteristics.\\
 We compare the proposed alternative scanning with the raster method in terms of damage behaviour on a beam sensitive commercial zeolite sample. Zeolites are considered a highly relevant class of microporous materials for their industrial applications such as in catalysis, ion exchanging, molecular sieving, etc. \cite{Davis2002}. Unfortunately atomic local analysis of its framework structure with (S)TEM has been difficult because of its poor stability under the electron beam \cite{Diaz2011,Ortalan2010}. Taking special care on the employed dose and/or employing a direct phase imaging technique, relatively recent works have already demonstrated the possibility to acquire atomic resolution STEM images before severe damage takes place \cite{Mayoral2011,Shen2020}. Although standard methods to synthesize these materials do exist \cite{H.Robson2001}, the use of self-prepared samples adds more variables to control in the experiments. Here, we deliberately opt for a commercially available sample as it enables others to repeat and reproduce our experiments and findings. This way we avoid the all-too-often circumstantial evidence that seems to surround the topic of beam damage in TEM and hinders progress in this important domain. \\
Using a freely programmable scan engine is another way how we increase the reproducibility of our findings by repeated automated experiments that rule out local variations in sample conditions. Our methodical experiments show that under the same conditions of total electron dose and dose rate, for both scanning methods, the alternative scanning systematically reduces electron beam damage. Our experimental findings show that beam damage is not only proportional to the total electron dose, but depends on the way the dose is distributed in time. In the second part of this manuscript we will attempt to empirically model this damage behaviour to agree qualitatively with all observations made in this experimental part, at least for this specific material. Empirical understanding gained from this toy model will give guidance on possible ways to further reduce beam damage in practical setups. We envision that this method can be applied to the wide family of zeolites and possibly beyond, to alleviate beam damage and allow for their structural characterization at the atomic scale. 

\section{Experimental}
\begin{figure}
\includegraphics[width=\columnwidth]{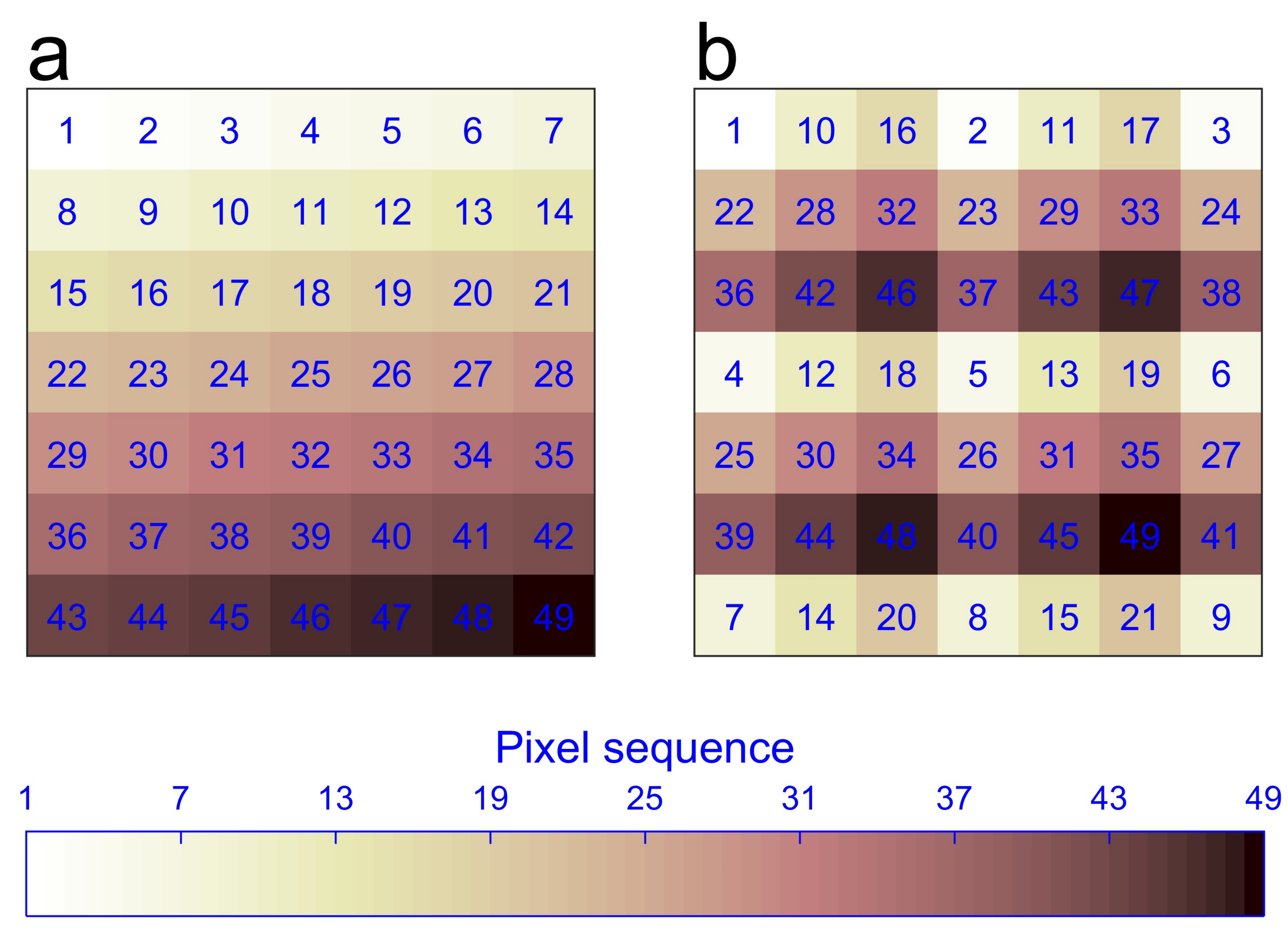}
\caption{\label{scansketch} Different scanning patterns for a 7 x 7 frame. The numbers and color scale indicate the order of the scanning positions. (a) Conventional raster scanning sequence. (b) Alternative scanning sequence, skipping two pixels in each scan direction. The total electron dose is the same in both cases; however, the temporal distribution of the dose is different. Note that the actual number of sampled points is far higher as displayed in this sketch, but the pattern is analogous.}
\end{figure}

Experiments were performed on a probe aberration-corrected FEI TITAN³ microscope, operated at 300~kV in HAADF-STEM mode, reaching a spatial resolution of less than 0.8~{\AA}  for a beam current of 50~pA (measured from a pico-ammeter connected to the flu-screen). The experiments were carried out on a commercial Linde Type A (LTA) zeolite sample (calcium exchanged sodium aluminium silicate, Sigma Aldrich BCR-705), which is among the most beam sensitive zeolites, Si/Al = 1, \cite{Mayoral2011}. The sample was crushed in a mortar for 5 min, dispersed in ethanol and drop casted on a holey carbon TEM grid.
A custom hardware scan engine \cite{Zobelli2020, commentouds} was employed to control the scan amplifier inputs of the microscope and to record the high-angle annular dark-field (HAADF) signal. A field-programmable gate array (FPGA) controls the synchronised feeding of the high speed digital to analog converters with a programmable pattern while sampling the input with an analog to digital converter. The recorded signal is then progressively displayed in a 2D array according to the scanning sequence described above. Figure \ref{scansketch} illustrates the scanning sequence for two different scanning methods. 

\begin{figure}
\includegraphics[width=\columnwidth]{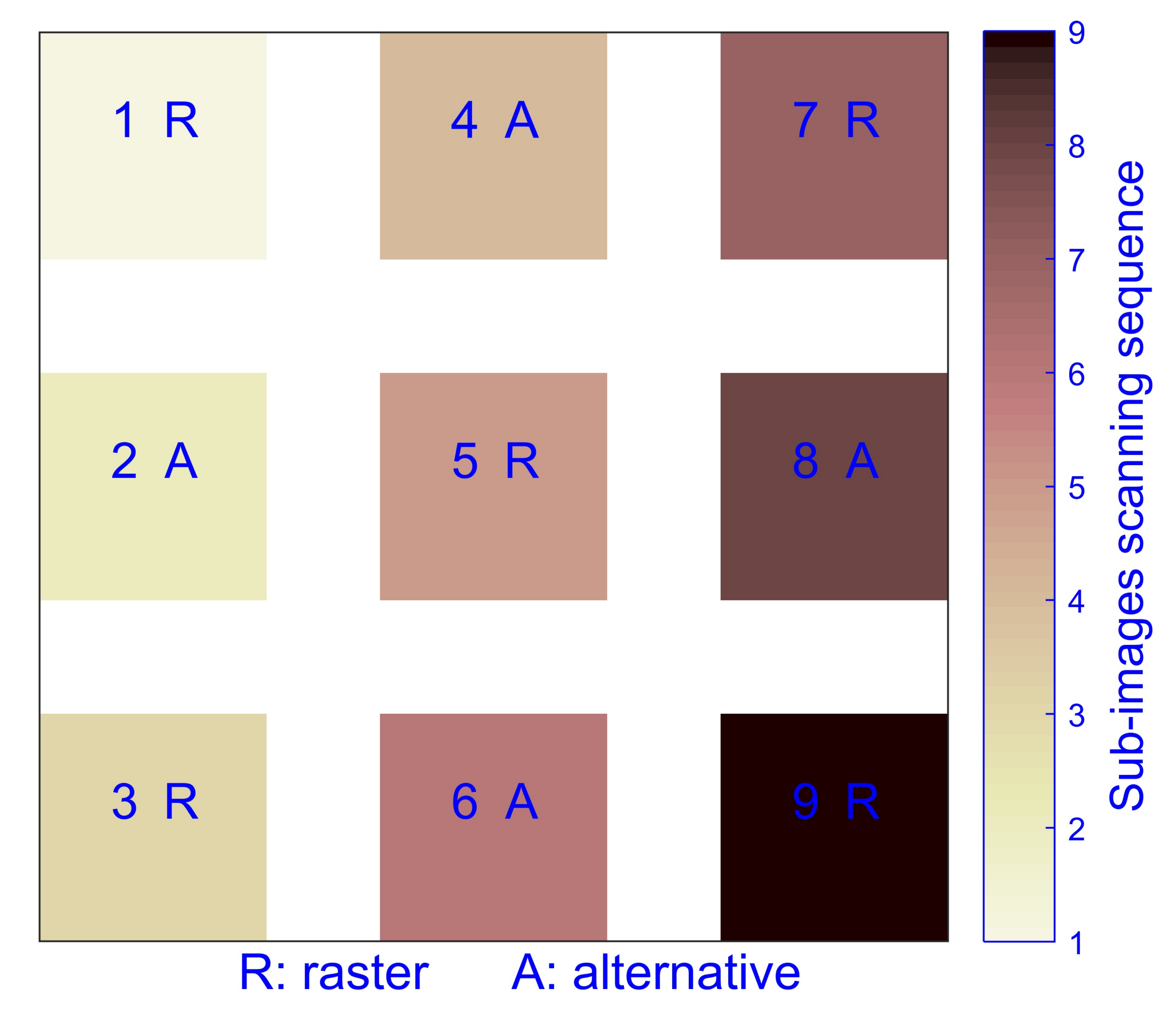}
\caption{\label{boxsketch}3 x 3 sub-images experiment performed on areas of interest of thin zeolite crystals. The numbers indicate the sequence of the squared scanned areas. The areas were scanned with either  the raster (R) or alternative (A) scanning pattern. The space between the scanned areas in the vertical and horizontal direction is half of the size of the scanned areas.}
\end{figure}

In order to have similar conditions for all the acquisitions, the raster and alternative scanning were compared by acquiring high resolution images on areas of thin crystals with uniform thickness showing the same [001] crystallographic orientation. Practically, the flexibility of the scan engine allowed us to acquire raster and alternative scans within the same image. As shown in Figure \ref{boxsketch}, single experiments consisted in scanning 3 x 3 sub-images continuously from top to bottom and from left to right, alternating the raster and alternative scanning. All the sub-images or scanned areas, consisting of a frame size of 512 x 512 pixels, were scanned with the same pixel size and dwell time. To avoid any pre-damage coming from earlier scanned sub-areas, these were spaced by a distance of half of the field of view of each area (256 pixels in the present case). A flyback delay, which corresponds to a pause in the beam movement before starting a new line, is commonly used to acquire raster images \cite{Velazco2020}. However, it is known to apply a considerable amount of dose on the left side of the scanned area, which contributes to a strong uneven distribution of the electron dose on the sample. In order to remove this extra source of damage, no flyback delay was employed for any of the scanning methods. To avoid any beam damage from the steady beam, first an area of interest was found scanning at low magnification, then the beam was blanked and the scanning was stopped. The experiments were performed at high magnification, allowing high resolution imaging, and the beam was manually unblanked immediately after the start of the first scanning sub-area and blanked before the scanning of the last sub-area finished. For this reason the first and last scanned sub-areas of the experiments, numbered 1 and 9 on Figure \ref{boxsketch}, were disregarded when comparing both scanning methods. Multiple acquisitions over the same area of interest were acquired following the same procedure, the time between consecutive acquisitions was approximately 6~s. Minor sample drift effects were observed during the acquisitions. These well-controlled experimental conditions allowed us to fairly compare the raster and alternative scanning methods at the same total electron dose and dose rate per pixel.

In this work the electron dose was calculated dividing the electron beam current, expressed in electrons per second, by the size of the individual scanned areas of the sub-images experiments and multiplying this value by the time it takes to scan that area. The instantaneous electron dose rate per pixel was calculated dividing the electron beam current, expressed in electrons per second, by the area of individual pixels. Since for each experiment, the same beam current, pixel size and dwell time were employed, the electron dose and electron dose rate per pixel were the same, independent of the scanning pattern. The experiments were performed at high resolution with two different pixel sizes, 24.3~pm and 34.3~pm, with corresponding dose rate per pixel of approximately $5.3 \times 10^9~\text{e}^-\text\AA{}^{-2}\text{s}^{-1}$ and $2.7 \times 10^9~\text{e}^-\text\AA{}^{-2}\text{s}^{-1}$, respectively.

\section{Results}
\begin{figure}
\includegraphics[width=\columnwidth]{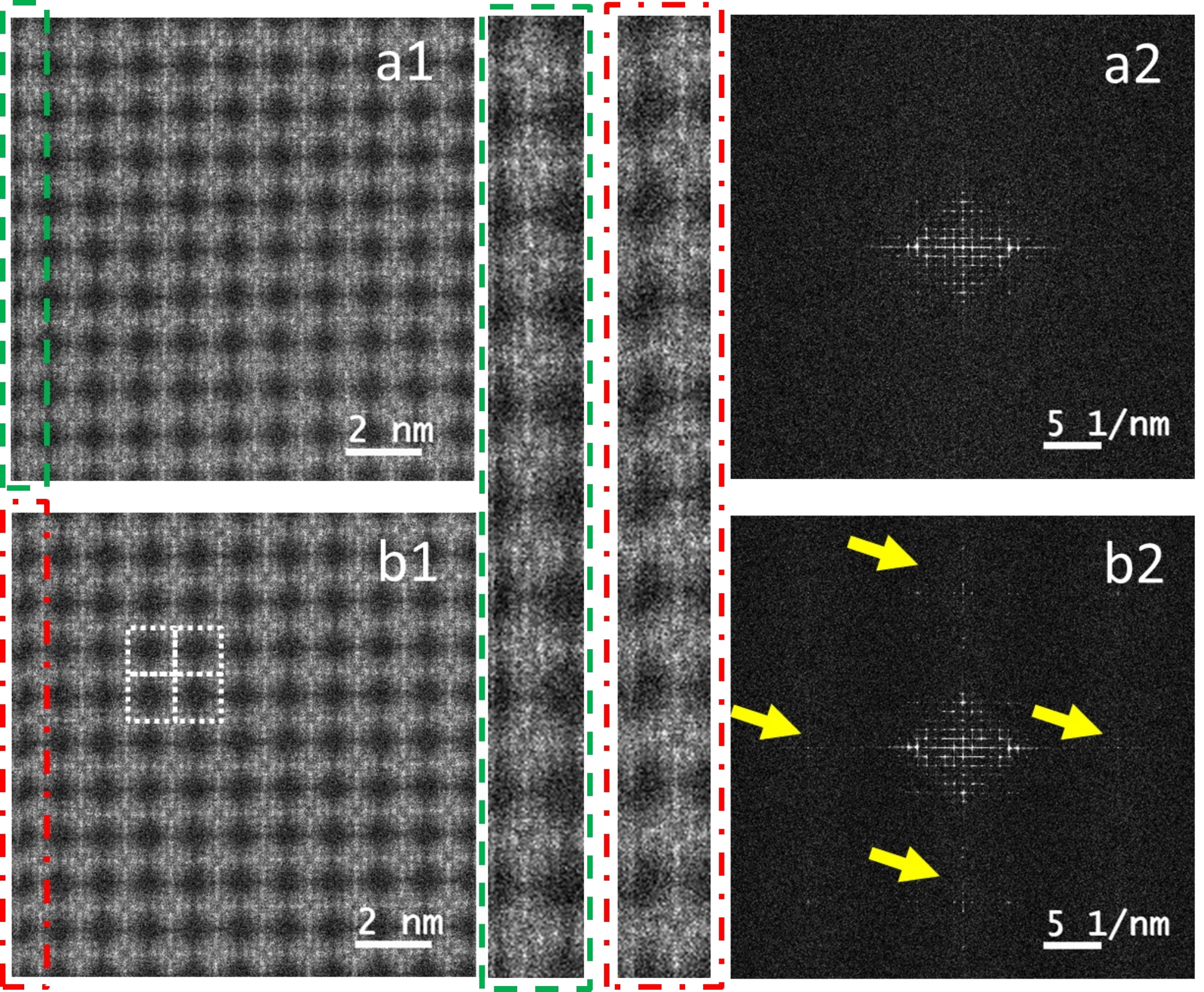}
\caption{\label{highrescomparison}Images extracted from a 3 x 3 sub-images experiment acquired with 24.3~pm pixel size, 6~$\mu\text{s}$ dwell time and a calculated dose of $3.17\times10^4~\text{e}^-\text\AA{}^{-2}$. (a1) Image acquired with the raster scanning. (b1) Image acquired with the alternative scanning. Highlighted areas show image distortions on the left side of the images because of the settling time effect of the probe when no flyback delay is applied. The contrast on the raw images were adjusted equally for feature enhancement. (a2) Diffractogram calculated from the raster image. (b2) Diffractogram calculated from the image acquired with the alternative scanning. The extra spots indicated by the arrows are the effect of the misaligning scanning lines in the x and y directions because of the time response of the probe when skipping pixels. }
\end{figure}

Figure \ref{highrescomparison} shows two images acquired respectively with the raster and alternative methods. The images were extracted from a 3 x 3 sub-image experiment acquired with 24.3~pm pixel size, 6~$\mu\text{s}$ dwell time and a total dose of $3.17\times10^4~\text{e}^-\text\AA{}^{-2}$. Both images are comparable in terms of contrast and resolution as can be seen directly from the LTA framework in the HAADF image and the diffractograms; however the diffractogram that corresponds to the image acquired with the alternative method presents some extra spots at high frequency (indicated by the yellow arrows in Figure 3(b2)). The extra spots are the result of sample drift which in combination with the nonconsecutive scanning leads to a periodic modulation which can be corrected as described in our previous work \cite{Velazco2020}. However, this was not applied here since it involves interpolation methods that would unavoidably change the intensity values of the acquired pixels and contrast of the images, hampering objective comparisons. The effects of settling time of the probe can also be identified as distortions on the left side of the images, as no flyback delay is applied, see enlarged images of Figures \ref{highrescomparison}(a1) and \ref{highrescomparison}(b1). The distortions are more pronounced for the alternative scanning. The dashed squares in Figure \ref{highrescomparison}(b1) overlay over 2 x 2 crystal unit cells, the large cages inside the unit cells correspond to the alpha cages and the small cage in the center corresponds to the sodalite cage \cite{Mayoral2011}.

\begin{figure}
\includegraphics[width=\columnwidth]{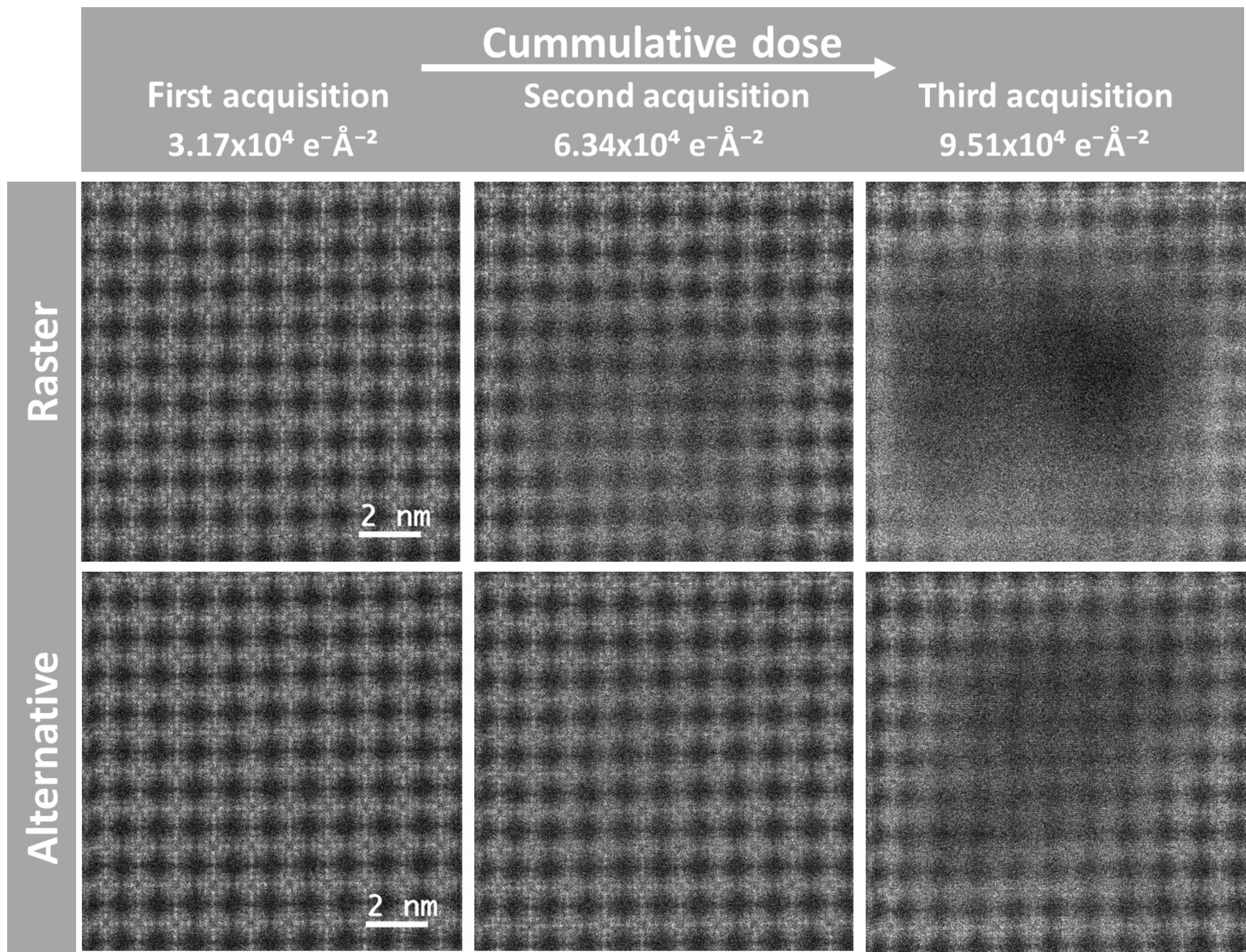}
\caption{\label{highresmultiple} Sub-images extracted from three consecutive acquisitions over the same 3 x 3 sub-image experiment. The scanning was performed with 24.3~pm pixel size, 6~$\mu\text{s}$ dwell time and a calculated dose of approximately $3.17\times10^4~\text{e}^-\text\AA{}^{-2}$ per acquisition. The contrast on the raw images is chosen  equal to allow a fair comparison of the evolution under beam damage.}
\end{figure}

\begin{figure}
\includegraphics[width=\columnwidth]{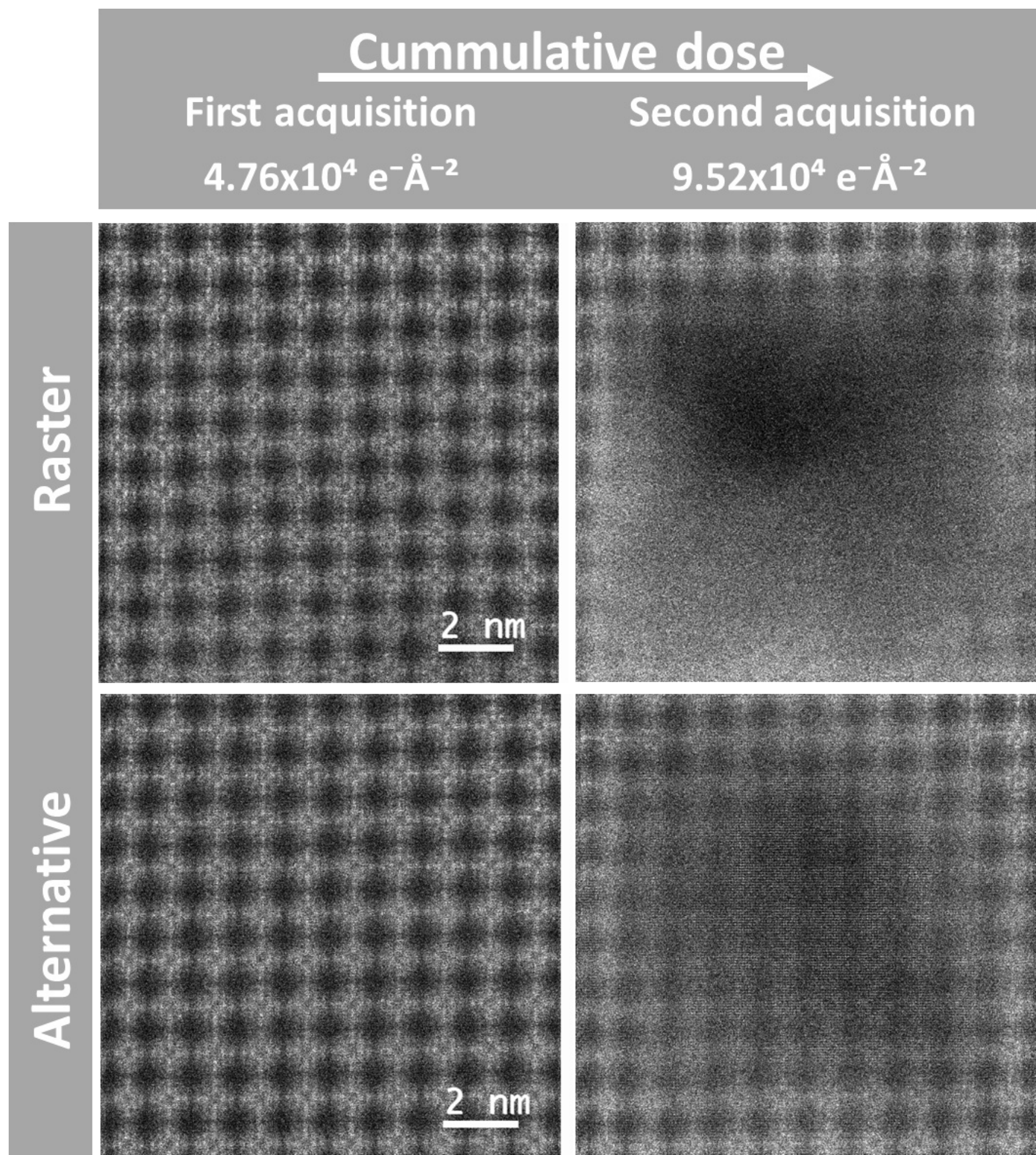}
\caption{\label{highresmultiple2} Sub-images extracted from two consecutive acquisitions over the same 3 x 3 sub-images experiment. The scanning was performed with 24.3~pm pixel size, 9~$\mu\text{s}$ dwell time and a calculated dose of approximately $4.76\times10^4~\text{e}^-\text\AA{}^{-2}$ per acquisition.  The contrast on the raw images is chosen  equal to allow a fair comparison of the evolution under beam damage.}
\end{figure}

Figures \ref{highresmultiple} and \ref{highresmultiple2} show multiple acquisitions extracted from two different 3 x 3 sub-images experiments performed on different areas of the same crystal. In both cases the scanning was performed with the same pixel size, 24.3~pm, and the same dose rate per pixel, $5.3\times10^9~\text{e}^-\text\AA{}^{-2}\text{s}^{-1}$. Each figure shows images acquired with the raster and alternative methods, the images correspond to the areas numbered 5 and 4, respectively, of the experiments (see Figure \ref{boxsketch}). Figure \ref{highresmultiple} corresponds to an experiment carried out scanning at 6~$\mu\text{s}$ dwell time with an electron dose of $3.17\times10^4~\text{e}^-\text\AA{}^{-2}$, three consecutive acquisitions were performed for this experiment. Figure \ref{highresmultiple2} corresponds to an experiment carried out scanning at 9~$\mu\text{s}$ dwell time with an electron dose of $4.76\times10^4~\text{e}^-\text\AA{}^{-2}$, two consecutive acquisitions were performed for this experiment. The total accumulated dose on each scanned area of both experiments is the same, $9.51\times10^4~\text{e}^-\text\AA{}^{-2}$.

As can be seen from  Figures \ref{highresmultiple} and \ref{highresmultiple2} , the HAADF signal changes progressively as the accumulated dose increases, which indicates sample degradation. Loss of mass and amorphization can be distinguished as areas that become dark and areas where the framework structure becomes blurred, respectively. These changes are more visible in the central-bottom region of the areas acquired with the raster method while mainly in the central region of the areas acquired with the alternative method. A clear observation in both experiments is that degradation of the sample is more pronounced when scanning with the raster method, except for the first acquisitions where the image quality seems to be comparable for both scan types. Moderate beam damage after the first acquisitions is observed, blurring of the sodalite cages to some extent. Deformation of the structure is apparent as a curved framework on the top region of the images for the last acquisitions acquired with both scanning methods. In general, after applying the total electron dose, the experiment performed at 9~$\mu\text{s}$ dwell time presents more damage than the experiment performed at 6~$\mu\text{s}$ dwell time, for both raster and alternative methods, despite the fact that the total accumulated dose is the same in all four cases.

\begin{figure}
\includegraphics[width=\columnwidth]{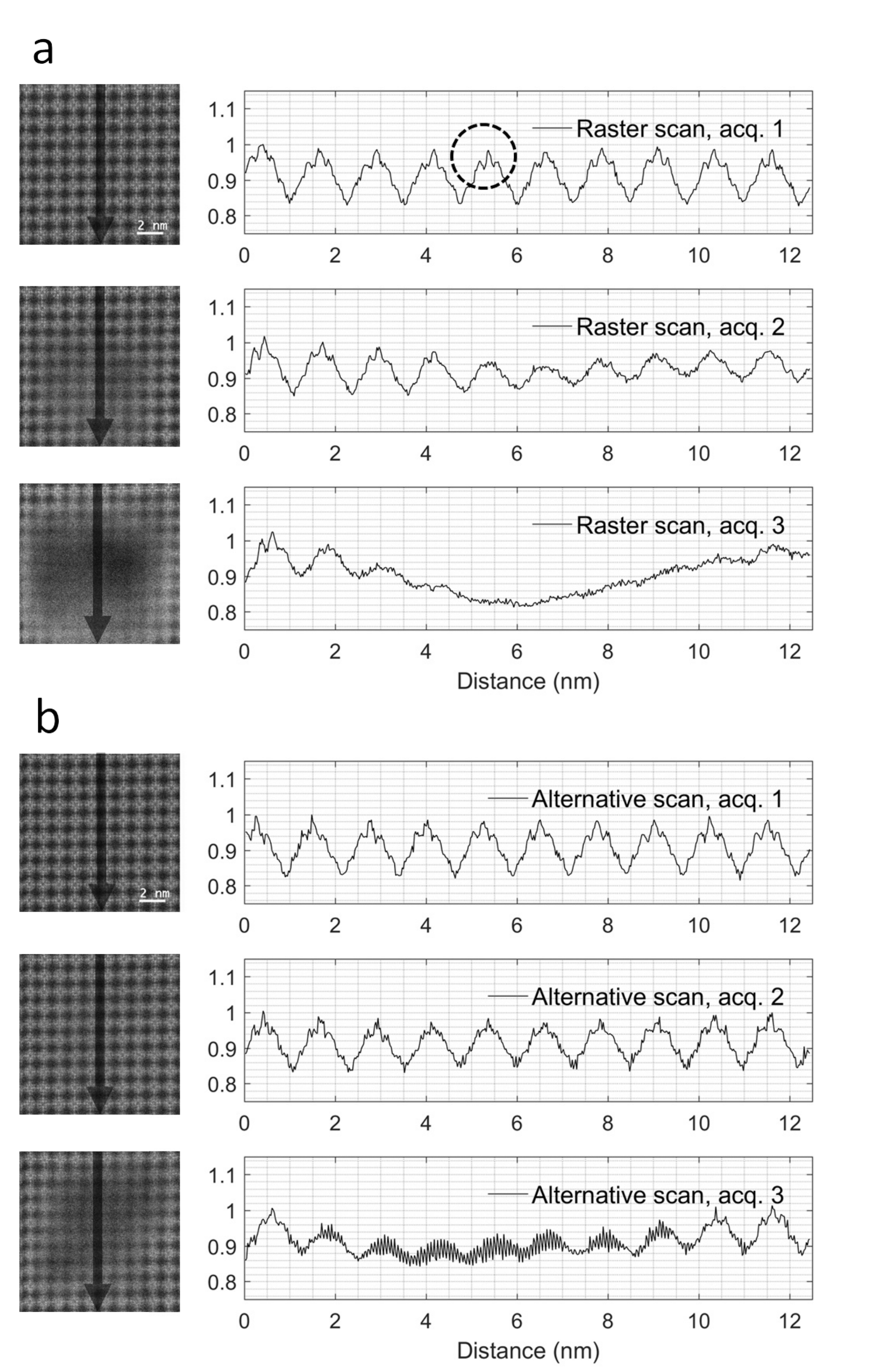}
\caption{\label{lineprofile} Integrated line profiles of images in Figure \ref{highresmultiple}, experiments performed at 6~$\mu\text{s}$ dwell time. Line profiles extracted from the center of the images with a 256 pixel integration width. Line profile corresponding to the experiment scanning with the raster method (a) and alternative method (b).}
\end{figure}

\begin{figure}
\includegraphics[width=\columnwidth]{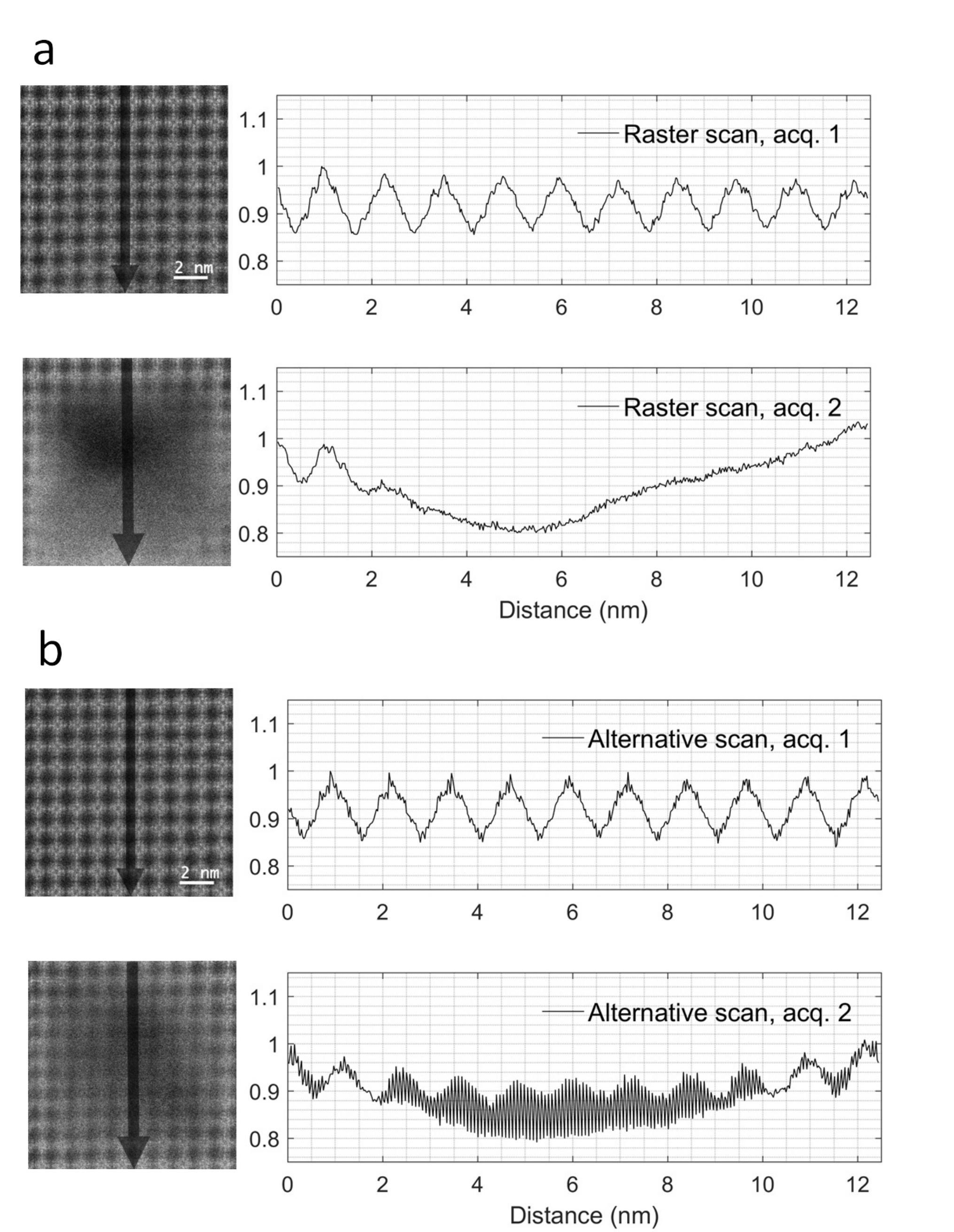}
\caption{\label{lineprofile2} Integrated line profiles of images in Figure \ref{highresmultiple2}, experiments performed at 9~$\mu\text{s}$ dwell time. Line profiles extracted from the center of the images with a 256 pixel integration width. Line profile corresponding to the raster method (a) and alternative method (b).}
\end{figure}

Figures \ref{lineprofile} and \ref{lineprofile2} show integrated line profiles from the images in Figures  \ref{highresmultiple} and \ref{highresmultiple2}, respectively. The profiles were extracted from the center of the images, where damage is most apparent, along the black arrows in the figures and considering a width of 256 pixels. The intensities were normalized with respect to the maximum intensity of the integrated line profile corresponding to the first acquisition. The line profiles in Figure \ref{lineprofile} correspond to the images acquired at 6~$\mu\text{s}$ dwell time, shown in Figure \ref{highresmultiple}. For the first acquisition, the subnanometer scale variations in the profile, small peak fluctuations inside the dashed circle in Figure 6(a), correspond to variations in the density of atomic columns around the sodalite cages. The larger variations, at nanometer scale, correspond to variations on the density of atomic columns around the unit cells of the framework. The line profiles corresponding to the first acquisition with both scanning methods are quite comparable. In Figure \ref{lineprofile}(a), second acquisition, the loss of the subnanometer scale variations and the reduced signal amplitude are clearly visible on the central region of the profile, and to a lesser extent on the right region. While the profile from the data acquired with the alternative scanning, Figure \ref{lineprofile}(b), second acquisition, becomes more noisy without showing a clear reduction of the signal amplitude. In both cases, deformation of the structure is also evident here which is indicated by the shift of the positions of the leftmost peaks with respect to their positions on the first acquisitions. For the third acquisition with the raster method, the nanometer scale variations are completely lost mostly from the center to the right of the profile (middle to down direction in the images), which indicates amorphization of the sample. In the central region, the reduced intensity suggests loss of mass. For the data acquired with the alternative method, the profile still preserves the nanometer scale variations with a reduced signal amplitude on the central region; however, with a subnanometer scale modulation suggesting alternating loss of mass resulting from the periodic array of pixels that came early in the scan with respect to those that came later. Here the shift of the peaks continues for the leftmost positions.

In Figure \ref{lineprofile2} the line profiles correspond to the images acquired at 9~$\mu\text{s}$ dwell time, displayed in Fig. \ref{highresmultiple2}. The line profile corresponding to the first acquisition acquired with the raster method, Figure \ref{lineprofile2}(a), already shows missing subnanometer scale variations mostly in the central region of the profile while the profile corresponding to the alternative method already looks noisy. For the second acquisition with the raster method, the nanometer scale variations are completely lost mostly from the center to the right of the profile, which indicates that amorphization of the sample is enhanced in the scanning direction. In the central region, the reduced intensity suggests loss of mass while the increased intensity to the right indicates mass accumulation. For the data acquired with the alternative method, the profile still preserves the nanometer scale variations with a clear reduced signal amplitude on the central region; however, with a subnanometer scale modulation suggesting alternating loss of mass.

\begin{figure}
\includegraphics[width=\columnwidth]{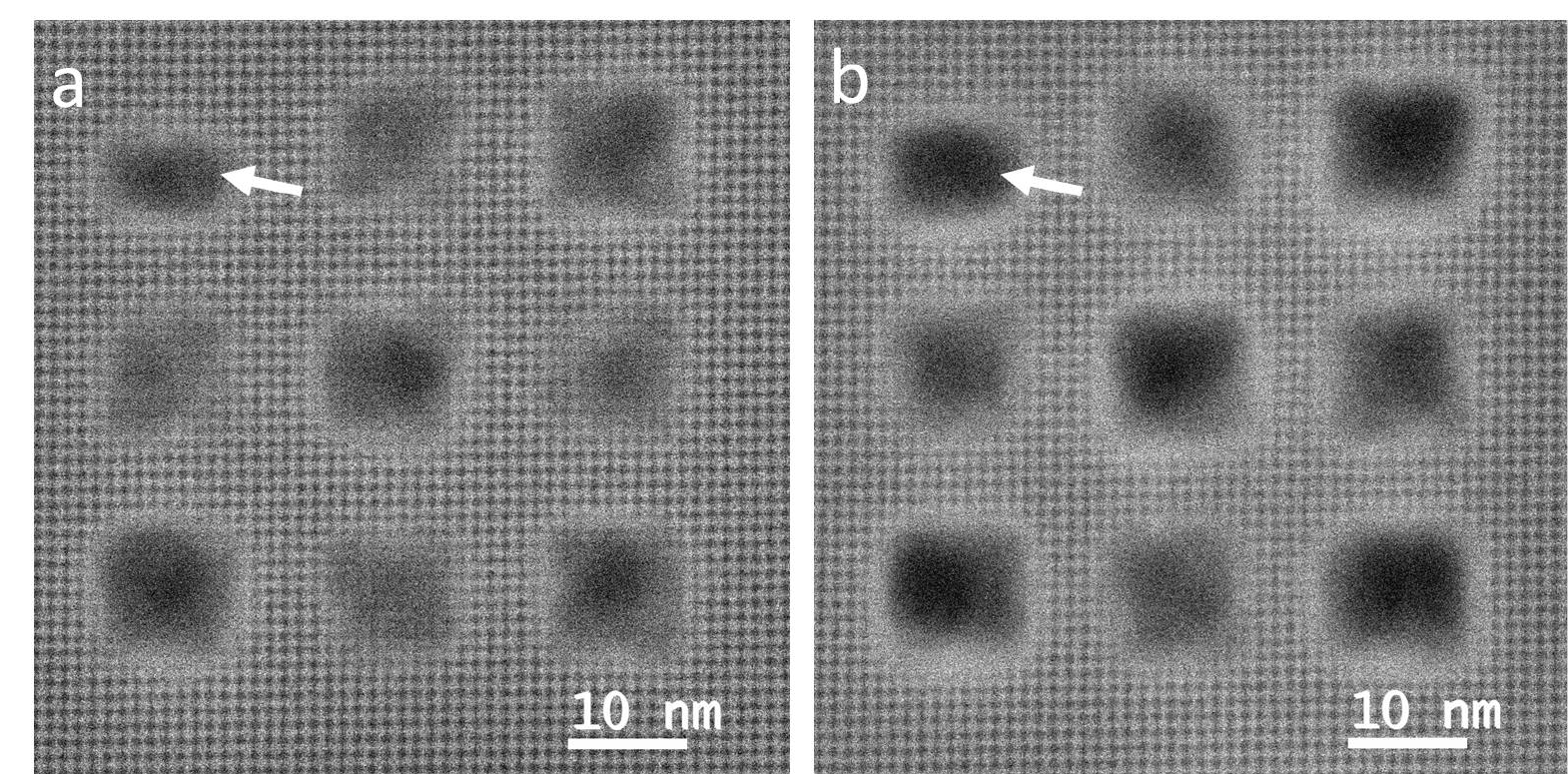}
\caption{\label{largefield} Large field of view acquisitions after performing 3 x 3 sub-images experiments presented in (a) Figure \ref{highresmultiple} (3 aquisitions at 6~$\mu\text{s}$) and (b) Figure \ref{highresmultiple2} (2 aquisitions at 9~$\mu\text{s}$) using identical contrast settings. As indicated by the contrast, the areas scanned with the alternative method show less damage in both experiments. When comparing areas scanned with the same method in both experiments, less damage is observed in (a) compared to (b) even though both obtained the same electron dose. The white arrow indicates the first subimage that is only illuminated partially due to the manual opening of the beam shutter and is ignored for further analysis.}
\end{figure}

The observations described above were consistent for all the scanned areas of the 3 x 3 sub-images experiments. Figure \ref{largefield} shows larger field of view acquisitions recorded after the experiments from which the images in Figure \ref{highresmultiple} and \ref{highresmultiple2} were extracted. The images were acquired with the conventional raster method at a lower magnification, at 3~$\mu\text{s}$ dwell time, after the last acquisition of the corresponding experiment finalized. The flyback delay was set to zero, and only cropped central regions, exempt from distortions, are shown here. The areas scanned with the raster method, areas numbered 1, 3, 5, 7 and 9 as depicted in Figure \ref{boxsketch}, are darker indicating an increased loss of mass compared to the areas scanned with the alternative method. The bright edges surrounding the scanned areas indicate accumulation of mass. The mass probably diffused from the scanned area (darker) to it surrounding (brighter) during the scanning. More damage is noticed when comparing the last acquisitions of the experiments in Figures \ref{highresmultiple} and \ref{highresmultiple2} with these large field of view images. The increased damage could be explained by its dynamic nature which would explain damage propagation even during intervals of no-irradiation and/or could be the effect of unavoidable beam damage during this large field acquisition itself, even at lower magnification and reduced dwell time. Although, in the course of the experiments, negligible beam damage was observed from single acquisitions at those conditions; however, this could be enhanced by the accumulation of damage from the previous experiments performed at much higher magnification. Here, the increased damage of the experiment acquired at 9~$\mu\text{s}$ dwell time compared to 6~$\mu\text{s}$ dwell time is also clear. For the 9~$\mu\text{s}$ dwell time experiment, the scanned areas, scanned with the raster and alternative methods, are darker than in the 6~$\mu\text{s}$ dwell time experiment, furthermore the edges are brighter.

\begin{figure}
\includegraphics[width=\columnwidth]{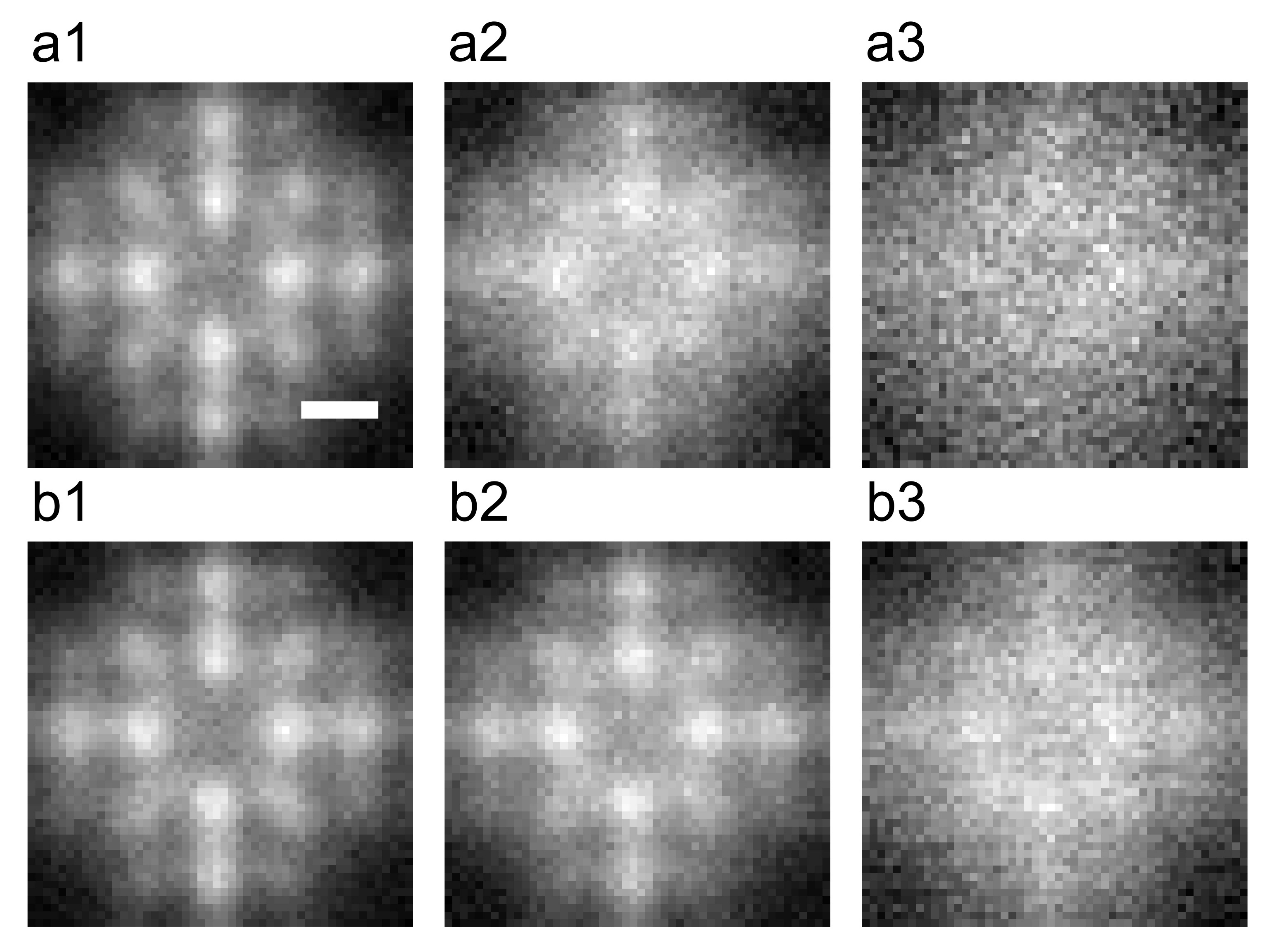}
\caption{\label{averaged} Averaged unit cell image of the sodalite cage obtained with a template-matching procedure applied to the images from Figure \ref{highresmultiple} (6~$\mu\text{s}$). (a1), (a2) and (a3), represent first, second and third acquisitions with the raster method, while (b1), (b2) and (b3) represent results from the alternative scan method. The number of cells found in each case is listed in Table \ref{ncells}. (scale bar 0.2~nm).}
\end{figure}

\begin{figure}
\includegraphics[width=\columnwidth]{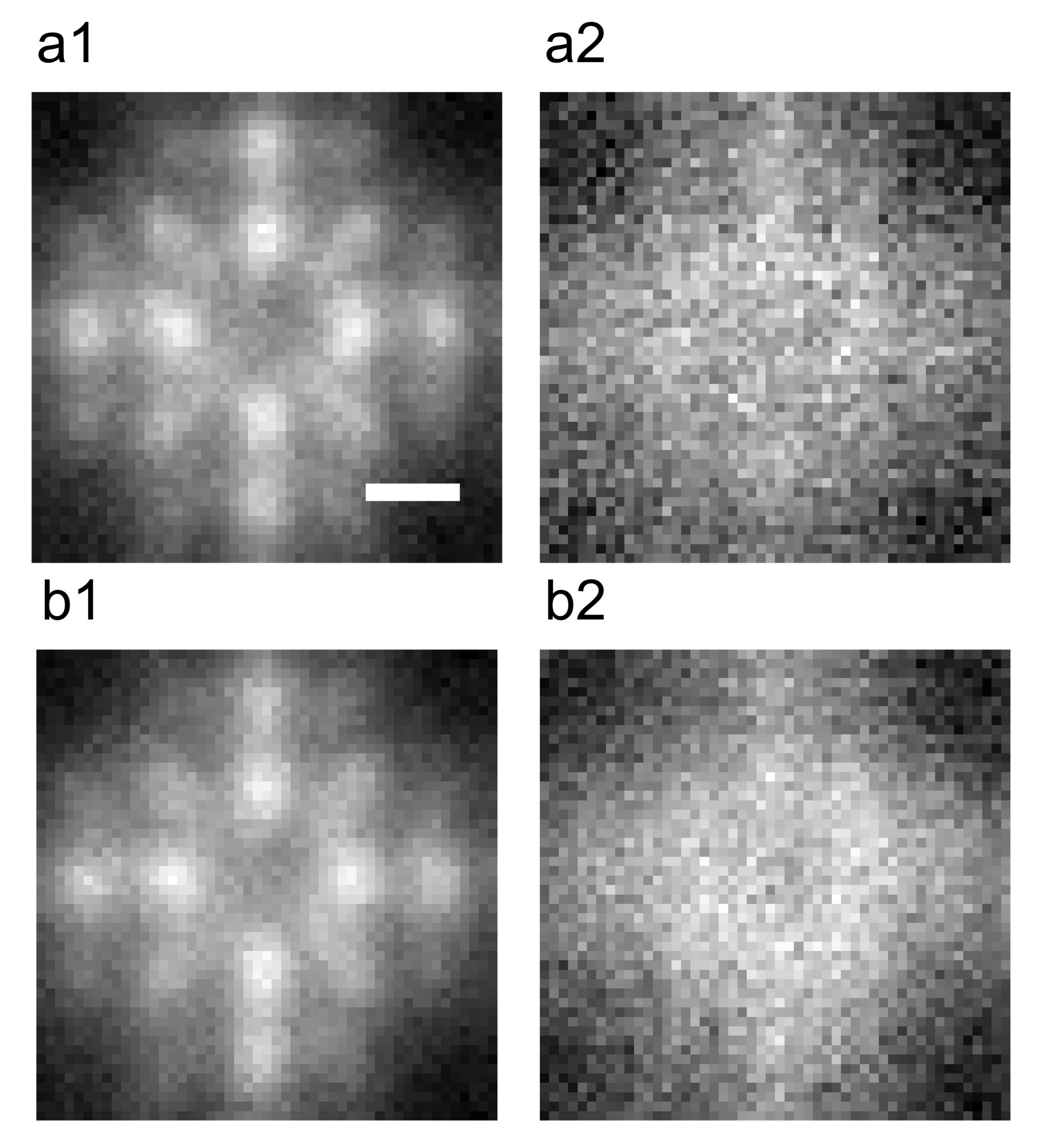}
\caption{\label{averaged2} Averaged unit cell image of the sodalite cage obtained with a template-matching procedure applied to the images from Figure \ref{highresmultiple2} (9~$\mu\text{s}$). (a1) and (a2), represent first and second acquisitions with the raster method, while (b1) and (b2), represent results from the alternative scan method. The number of cells found in each case is listed in Table \ref{ncells}. (scale bar 0.2~nm).}
\end{figure}

Beam damage can be quantified by the local information at the unit cell scale and can be used to monitor the information content as a function of the applied dose. A template-matching procedure based on a cross-correlation function was employed to find regions (unit cells) of the images that have similar features compared to a template \cite{PRZYBYA2012}. These similarities were evaluated as a function of the applied dose. The procedure was  performed over the first acquisition of the experiments, which exhibits moderate damage. A template of 50 x 50 pixels size corresponding to the sodalite cage was selected and an averaged image was obtained from all the cells that showed a cross correlation $\geq$ 45\% of the maximum obtained cross correlation value with respect to the sum of all previously selected unit cells. This averaged image was used iteratively as a new template for all the acquisitions. First, the template-matching methodology was applied to a Gaussian (sigma = 3.5 pixels) filtered version of the raw images, which exhibit low SNR and poor contrast because of beam damage effects. Then, the positions of the cells that matched the templates were used to extract the cells from a new template-matching procedure applied to the raw unfiltered images. Because of the distortions on the left side of the images, the first left column of the sodalite cages was excluded during the template-matching procedure. The first or last row of the sodalite cages was excluded only in cases where the cages were not fully imaged on the first acquisitions. The same procedure was applied separately for each of the scanned areas in the 3 x 3 sub-images experiments. With this procedure, an averaged template image can be also obtained for all the further acquisitions of the experiments. Figures \ref{averaged} and \ref{averaged2} show the averaged images of the sodalite cage obtained from each of the acquisitions on Figures \ref{highresmultiple} and \ref{highresmultiple2}, experiments performed at 6 and 9~$\mu\text{s}$ dwell time, respectively. In both figures, reduced loss of the structure is evident after the first acquisitions done with the alternative method. Table \ref{ncells} summarizes the number of cells found with this procedure.  

%%\table1
\begin{table}
\centering
\caption{\label{ncells}Number of cells averaged in the template-matching procedure.}
\label{template cells}
\resizebox{\columnwidth}{!}{%
\begin{tabular}{ccccc} 
\cline{2-5} %%draw line from  column 2 to 5
 & \multicolumn{2}{c}{6~$\mu\text{s}$ dwell time} & \multicolumn{2}{c}{9~$\mu\text{s}$ dwell time}\\
 & \multicolumn{2}{c}{Number of cells} & \multicolumn{2}{c}{Number of cells}\\
\cline{2-5}
 & Raster & Alternative & Raster & Alternative\\
 & scanning & scanning & scanning & scanning\\
\hline
1st acquisition & 81 & 81 & 81 & 81\\
2nd acquisition & 81 & 81 & 35 & 81\\
3rd acquisition & 55 & 81 &  & \\
\hline\\
\end{tabular}
}
\end{table}

The reduced number of cells found in the last acquisitions for the raster method is the result of severe loss of mass and amorphization in the central-bottom regions making it difficult to recognise the template. In these cases the cells were found mainly on the edges of the images. Supporting Information Figure s1 shows the location of the cells on the three acquisitions of the raster images from Figure \ref{highresmultiple}. Their locations with respect to the first acquisitions may change because of sample deformation or drift of the sample. As indicated earlier, deformation of the structure was observed predominantly on the top region of the images while negligible drift was noticed.

\begin{figure}
\includegraphics[width=\columnwidth]{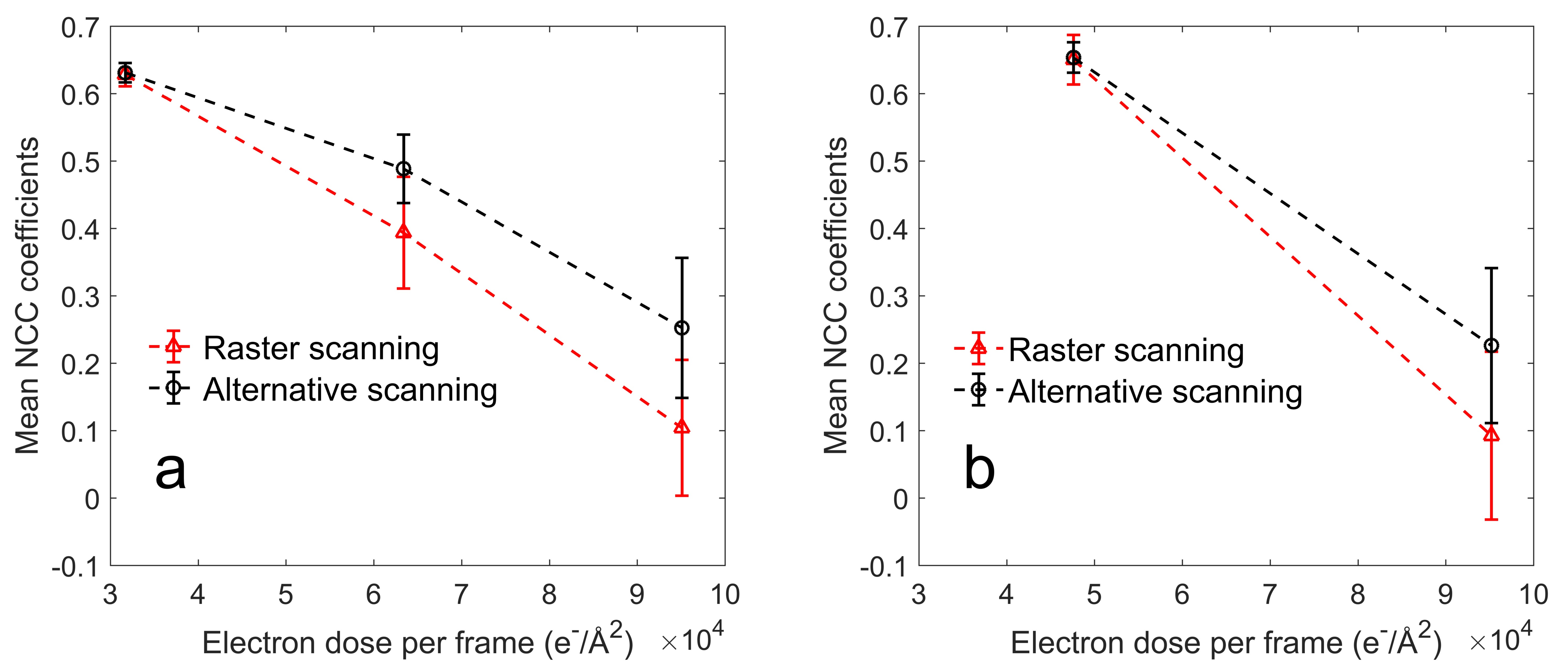}
\caption{\label{meanncc} Mean NCC coefficients, obtained with template-matching as a function of  accumulated dose per frame. The error bars pertain to the standard deviation over the multiple areas scanned in each experiment. (a) Plot corresponding to the three acquisitions of the experiment performed at 6~$\mu\text{s}$ dwell time. (b) Plot corresponding to the two acquisitions of the experiment performed at 9~$\mu\text{s}$ dwell time.}
\end{figure}

As the sample degradation increases, the cells representing the sodalite cages deviate more from the ideal averaged image. The degree of similarity can be represented by the normalized cross-correlation (NCC) coefficients and can serve as a quantative parameter representing beam damage \cite{Leijten2017}. Sample deformation, amorphization, contrast loss, increased noise, etc., all reduce the NCC. 
Here, the indices corresponding to the NCC coefficients of all the cells found on the first acquisitions were used to extract the coefficients for the next acquisitions. This way the same number of cells are always considered, 81 cells in all cases, avoinding the underestimation of beam damage from unit cells which are no longer recognisable. The mean of the NCC coefficients was calculated for all the scanned areas of the 3 x 3 sub-images experiments and plotted as a function of the applied dose in Figure \ref{meanncc}. 

With both scan patterns the mean of the NCC coefficients decreases as a function of dose. The mean of the coefficients of the first acquisitions is quite similar for the raster and alternative scanning. However, for the further acquisitions (as the dose increments), the average NCC for raster scanning is significantly lower when compared to alternative scanning for the same collected dose. This agrees with the qualitative observations of beam damage in the images in Figures \ref{highresmultiple} and \ref{highresmultiple2} and in the line profiles in Figures \ref{lineprofile} and \ref{lineprofile2}, where damage is more pronounced when scanning with the raster method but now takes into account all scanned images for improved statistics and provides an error bar as the standard deviation across all unit cells.

The mean NCC coefficients for both experiments, from Figures \ref{meanncc}(a) and \ref{meanncc}(b) are given in Table \ref{NCCcoeff}. The reason for having similar coefficients for the first acquisitions of both experiments is due to the fact that the scanned areas from each experiment were compared to its own averaged image. Even though, as highlighted from the line profiles, the first acquisition of the experiment acquired at 9~$\mu\text{s}$ dwell time, applied dose $4.76\times10^4~\text{e}^-\text{\AA}{}^{-2}$, already shows more damage compared to the first acquisition of the experiment acquired at 6~$\mu\text{s}$ dwell time, applied dose $3.17\times10^4~\text{e}^-\text{\AA}{}^{-2}$. 
The coefficients are clearly reduced with respect to the first acquisitions. From Figure \ref{meanncc}(a), when scanning with the raster method, a reduction of 37.3\% and 83.4\% for the second and third acquisitions are observed, respectively. While for the alternative method, 22.5\% and 60.0\% of reduction are remarked for the second and third acquisitions, respectively. From Figure \ref{meanncc}(b), a reduction of 85.8\% and a reduction of 65.4\% for the second acquisitions scanning with the raster and alternative method are observed, respectively. For all the cases, the reduction of the coefficients obtained for the raster method are larger than the ones obtained for the alternative method.

%%Table 2  
\begin{table}

\centering
\caption{Mean NCC coefficients from Figures \ref{meanncc}(a) and \ref{meanncc}(b) obtained with the template-matching procedure.}
\label{NCCcoeff}
\resizebox{\columnwidth}{!}{%
\begin{tabular}{ccccc} 

\cline{2-5} %%draw line from  column 2 to 5
 & \multicolumn{2}{c}{6~$\mu\text{s}$ dwell time} & \multicolumn{2}{c}{9~$\mu\text{s}$ dwell time}\\
 & \multicolumn{2}{c}{Mean NCC coeficcients} & \multicolumn{2}{c}{Mean NCC coeficcients}\\
\cline{2-5}
 & Raster & Alternative & Raster & Alternative\\
 & scanning & scanning & scanning & scanning\\
\hline
1st acquisition & 0.628 & 0.631 & 0.650 & 0.653\\
2nd acquisition & 0.394 & 0.489 & 0.092 & 0.226\\
3rd acquisition & 0.104 & 0.252 &  & \\
\hline\\
\end{tabular}
}
\end{table}

\begin{figure}
\includegraphics[width=\columnwidth]{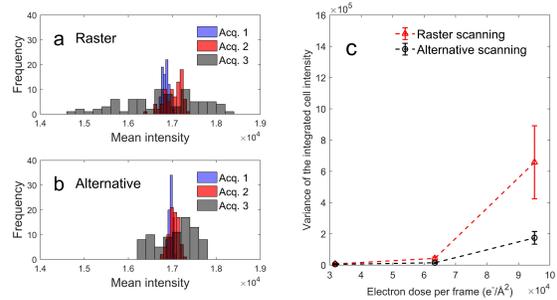}
\caption{\label{histogram} Histogram distribution and variance of the mean intensities of the cells found with the template-matching procedure.The locations of the cells found on the first acquisitions are kept for all the further acquisitions, and the mean intensities of each of the cells are calculated for all the acquisitions. Distributions corresponding to the experiments acquired at 6~$\mu\text{s}$ dwell time, 24.3~pm pixel size. (a) and (b) show the results for the raster and alternative scanning acquisitions, respectively, from Figure \ref{highresmultiple}. (c) Mean variance of the mean intensities as a function of the accumulated dose per frame. The mean of the variance is considered from the scanned areas of the 3 x 3 sub-images experiment. Areas numbered 3, 5 and 7 were considered for the raster scanning and areas numbered 2, 4, 6 and 8 were considered for the alternative scanning. The errors bars correspond to the standard deviation.}
\end{figure}

Another way to quantify the changes in the sodalite cages/cells is by the mean intensity in each cell. As the HAADF signal is proportional to the thickness of the sample, the changes in the signal corresponding to the sodalite cages can be interpreted as a change in density of the material, e.g. due to mass loss or accumulation of mass. Again, the locations of the cells found on the first acquisitions were kept for all further acquisitions to make sure also the more damaged areas are maintained in the evaluation. In order to maintain the absolute intensity, this methodology was applied to the raw images with an intensity scale zero calibrated with a blanked beam. The histogram distributions of the mean intensities of the cells corresponding to the first, second and third acquisitions from Figure \ref{highresmultiple} at 6~$\mu\text{s}$ dwell time, can be found in Figure \ref{histogram}. For the first acquisitions, low variation on the mean intensities of the cells is expected as the images were acquired on regions of uniform thickness, this is indeed observed as a narrow distribution in the histograms. However, as the applied dose increases, the histogram distribution widens as an indication of mass variation. For the second acquisition, with the raster scanning, a left-tailed distribution with the mean slightly shifted to the right is shown, which may suggest that mass is moving around from regions of the sample that become thinner. In Figure \ref{lineprofile}(a), the clear reduction of the signal amplitude observed in the central region of the line profile can be interpreted as thinning of the sample. For the alternative scanning, the distribution becomes wider, suggesting that the mass is re-distributed in the frame. The line profile in Figure \ref{lineprofile}(b) does not reveal any evident reduction of the signal amplitude. For the third acquisition, with the raster scanning, the histogram shows a wide distribution with tails towards lower intensities suggesting loss of mass. Likewise, for the alternative scanning, the distribution widens and skews even though the width remains lower than for the raster case and remains centered with respect to the mean of the first acquisition.  Figure \ref{histogram}(c) shows the mean of the variance from the histograms of the raster and alternative acquisitions calculated from the different scanned areas of the 3 x 3 sub-images experiment. The mean of the variance was calculated for the three acquisitions of the experiment, the error bars on the plot correspond to the standard deviation. For both scanning methods, the variance increases as a function of the dose. This is an indication of the broadening of the distributions which relates to the mass variations in the cells. Here the variances for the first acquisitions are quite similar for both methods. However, for the further acquisitions (as the dose is incremented), the values from the raster scanning show a significantly larger variance as compared to the alternative scanning. For the raster method, the variance of the second acquisition is 6 times the variance of the first acquisition. For the third acquisition, it is 94 times that value. For the alternative method, the variance of the average intensity in a unit cell of the second acquisition is 3 times the variance of the first acquisition. For the third acquisition, it is 37 times that value.

\begin{figure}
\includegraphics[width=\columnwidth]{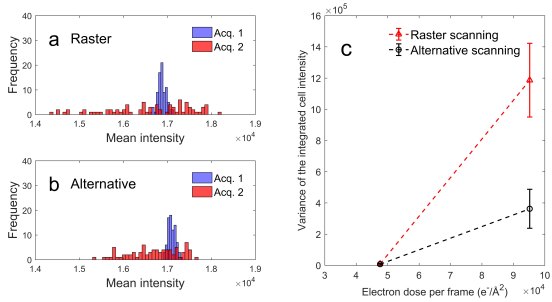}
\caption{\label{histogram2} Histogram distribution and variance of the mean intensities of the cells found with the template-matching procedure.The locations of the cells found on the first acquisitions are kept for all the further acquisitions, and the mean intensities of each of the cells are calculated for all the acquisitions. Distributions corresponding to the experiments acquired at 9~$\mu\text{s}$ dwell time, 24.3~pm pixel size. (a) and (b) show the results for the raster and alternative scanning acquisitions, respectively, from Figure \ref{highresmultiple2}. (c) Mean variance of the mean intensities as a function of the accumulated dose per frame. The mean of the variance is considered from the scanned areas of the 3 x 3 sub-images experiment. Areas numbered 3, 5 and 7 were considered for the raster scanning and areas numbered 2, 4, 6 and 8 were considered for the alternative scanning. The errors bars correspond to the standard deviation.}
\end{figure}

For the experiment acquired at 9~$\mu\text{s}$ dwell time, the histogram distributions of the mean intensities of the cells can be found in Figure \ref{histogram2}. Again, for the first acquisitions, narrow normal distributions are shown. For the second acquisitions, non-normal distributions are depicted. In both cases, the distributions are centered to the left with respect to the mean of the first acquisitions; however, the width of the distribution from the raster scanning is larger compared to the one from the alternative scanning. Similar to the previous experiment, this suggests loss of mass, being more severe for the raster scanning method. Figure \ref{histogram2}(c) shows the mean of the variance from the histograms of the raster and alternative acquisitions calculated from the different scanned areas of the 3 x 3 sub-images experiment. The mean of the variance was calculated for the two acquisitions of the experiment, the error bars on the plot correspond to the standard deviation. For both scanning methods, the mean of the variance increases as a function of the dose. The variances for the first acquisitions are quite similar for both methods. However, for the second acquisitions, the values from the raster scanning deviates from the value of the alternative scanning. For the raster method, the variance of the second acquisition is 132 times the variance of the first acquisition. For the alternative method, the variance of the second acquisition is 48 times the variance of the first acquisition. 
Loss of mass is apparent from the histogram of the experiment at 6~$\mu\text{s}$ dwell time scanning with the raster method, while for the experiment performed with the alternative method at 6~$\mu\text{s}$ dwell time, only a re-distribution of the mass is proposed from the histogram. Despite the fact that the total accumulated dose in both experiments is the same (experiments consisting in three acquisitions at 6~$\mu\text{s}$ dwell time and two acquisitions at 9~$\mu\text{s}$ dwell time), increased loss of mass is apparent from the last histograms of each of the experiments at 9~$\mu\text{s}$ dwell time. These findings suggest reduced damage scanning with the alternative method and for distributing the dose in more acquisitions scanning at a higher speed. This can also be interpreted from the plots on Figures \ref{histogram}(c) and \ref{histogram2}(c) where the variance from the last acquisitions is less when scanning at those conditions.

From the large field of view images acquired after performing the 3 x 3 sub-images experiments, images in Figure \ref{largefield}, relative mass loss can be quantified by calculating the ratio between the mean HAADF intensities of the scanned areas and the non-scanned areas in the experiments. Negligible beam damage was considered when acquiring the large field of view images. The relative quantification with respect to the non-scanned areas will allow us to compare the amount of damage between different block experiments. Only the central region, half of the size of the scanned areas, were considered for the calculations as the damage was observed to be more uniform there. The relative values were calculated with respect to the mean HAADF intensity of the areas close to the corners of the large field of view images in Figure \ref{largefield}, non-scanned regions during the experiments, and are tabulated in  Table \ref{mean HAADF}.  We deduce two clear trends from this table:
\begin{itemize}
\item{alternative scanning suffers significantly less mass loss when compared to raster scanning (7.7\% and 11\% for the 3 and 2 acquisition experiments, respectively, using the same total dose)}
\item{3 acquisitions at 6~$\mu\text{s}$ dwell time cause significantly less mass loss as compared to 2 acquisitions with 9~$\mu\text{s}$ dwell time for both scan patterns even though they represent the same total dose and total recording time (12.5\% for raster, 9.2\% for alternative).}
\end{itemize}

%%Table 3
\begin{table}

\centering
\caption{Relative mean HAADF intensity between the scanned areas and non-scanned areas of the two different 3 x 3 sub-images experiments shown by the large field of view images on Figure \ref{largefield}.}
\label{mean HAADF}
\resizebox{\columnwidth}{!}{%
\begin{tabular}{ccccc}  

\cline{2-5} %%draw line from  column 2 to 5
 & \multicolumn{2}{c}{6~$\mu\text{s}$ dwell time} & \multicolumn{2}{c}{9~$\mu\text{s}$ dwell time}\\
 & \multicolumn{2}{c}{3 acquisitions} & \multicolumn{2}{c}{2 acquisitions}\\
\cline{2-5}
 & Raster & Alternative & Raster & Alternative\\
 & scanning & scanning & scanning & scanning\\
\hline
Relative HAADF & $86.5 \pm~0.1$ & $94.2 \pm~0.1$ & $74.0 \pm~0.1$ & $85.0 \pm~0.1$\\
intensity (\%) &  & & & \\

\hline\\
\end{tabular}
}
\end{table}

More experiments were conducted over the same crystal at different dwell times and pixel size, showing similar results; less damage was observed on areas scanned with the alternative method. Images corresponding to an experiment acquired at 12~$\mu\text{s}$ dwell time and 34.3~pm pixel size can be found in the Supporting Information Figure s2. 

According to the assumed role of diffusion in the damage process (see part II of this manuscript), an area that will be scanned at some point in the acquisition sequence could be already affected while scanning the other areas. For example, the central region scanned with the raster method, area numbered 5 in Figure \ref{boxsketch}, which is enclosed by the rest of the neighboring areas, would potentially suffer more. To clarify whether or not the scanning sequence has some influence on the damage patterning, acquisitions were conducted inverting the sequence of the scanning methods. Experiments were performed starting with alternative scanning instead of raster scanning and alternating the methods afterward. The results reported previously were replicated in a different crystal with the inverted scanning sequence; more damage was found in the areas scanned with the raster method, independent of the scanning sequence. A large field of view image acquired after the sub-images experiment is shown in the Supporting Information Figure s3. 

\begin{figure}
\includegraphics[width=\columnwidth]{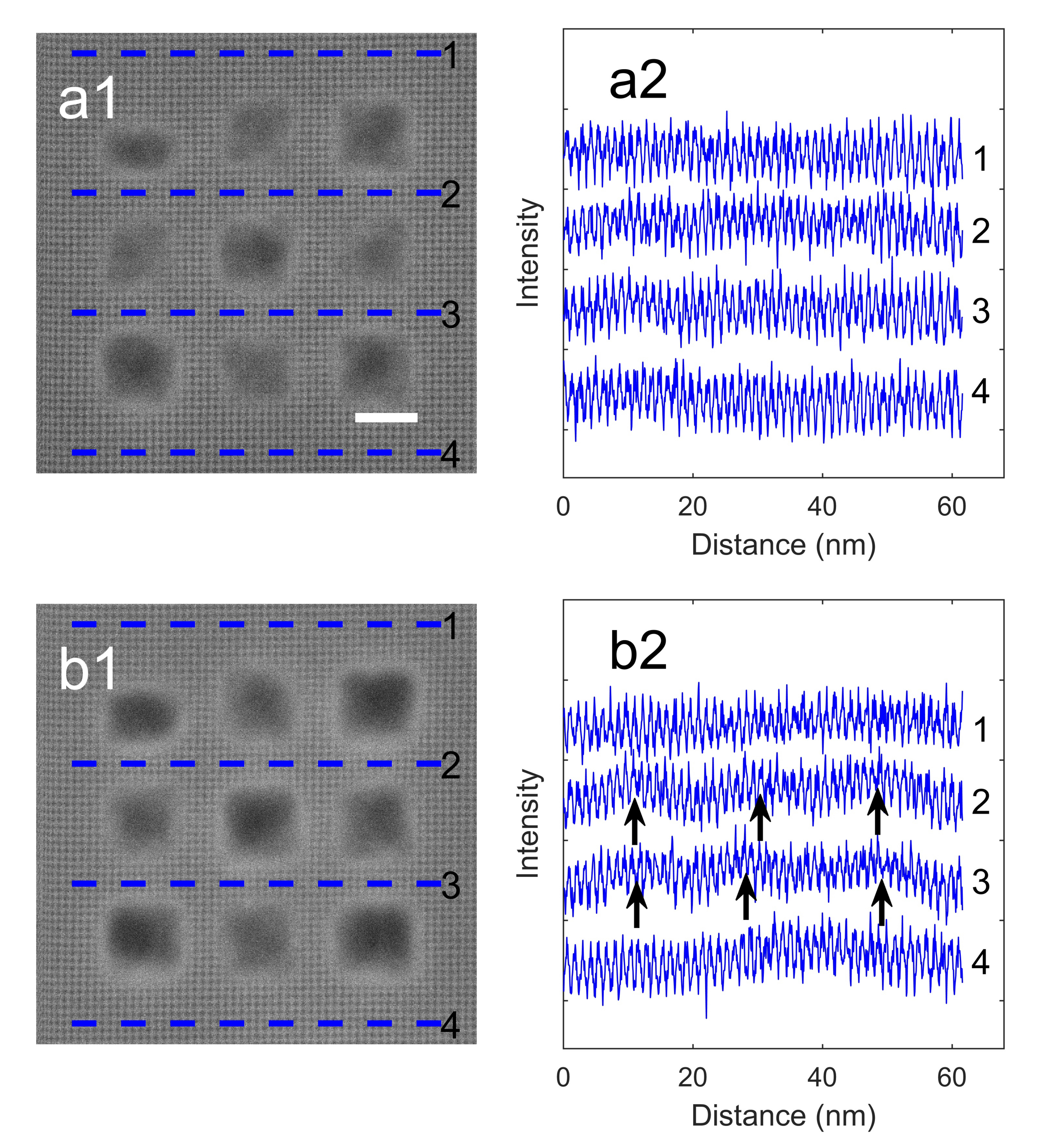}
\caption{\label{inbetweenbox} Integrated line profiles, with 15 pixels width, over the large field of view images acquired after performing the 3 x 3 sub-images experiments shown in Figure \ref{largefield}. (a1) and (b1) correspond to the experiments acquired at 6~$\mu\text{s}$, three acquisitions, and 9~$\mu\text{s}$ dwell time, two acquisitions, respectively, with 24.3~pm pixel size. (a2) and (b2) show the data from the line profiles on (a1) and (b1), respectively. To clearly compare the profiles, a fixed offset was added to the intensities of line 1 and line 2, and subtracted to the data of line 4. The mean intensity and amplitude of the raw data from the different line profiles is quite comparable. Besides the intensity variations because of the framework structure, the line profiles in (a2) are mostly flat. Similar characteristics can be seen from the profiles of line 1 and 4 in (b2); however, lines 2 and 3 show three bumps, indicated by arrows, at the same positions where the areas in (b1) were scanned. The scale bar represents 10~nm.}
\end{figure}

Although no apparent beam damage was found inbetween the scanned areas of the experiments, for the acquisitions at 9~$\mu\text{s}$ dwell time, some accumulation of mass was observed in those regions as an increased intensity. The gap between the sub-images was approximately 6.2~nm distance, the scanning was performed with 24.3~pm pixel size.  As mentioned previously, the mass removed from the central regions was accumulated mainly in the surroundings and could probably diffuse beyond as a result of the experiment itself or when acquiring the large field of view images shown in Figure \ref{largefield}. Figure \ref{inbetweenbox} shows integrated line profiles, with 15 pixels width, taken horizontally over the large field of view acquisitions of the experiments performed at 6~$\mu\text{s}$, three acquisitions, Figure \ref{inbetweenbox}(a1), and 9~$\mu\text{s}$ dwell time, two acquisitions, Figure \ref{inbetweenbox}(b1). Line 1 and line 4, correspond to the top and bottom regions, respectively, enclosing the area of the experiments, line 2 and line 3 correspond to the regions in between the actual scanned areas. A fixed offset was added to the intensities from the data of line 1 and line 2, and subtracted to the data of line 4 in Figures \ref{inbetweenbox}(a2) and \ref{inbetweenbox}(b2) to clearly compare the profiles. The mean intensities and amplitudes of the raw data from the different line profiles are comparable, the experiments were done over the same crystal in areas of uniform thickness. Besides the intensity variations because of the framework structure, the line profiles in Figures \ref{inbetweenbox}(a2) are mostly flat. Similar characteristics can be seen from the profiles of line 1 and 4 in Figure \ref{inbetweenbox}(b2); however, lines 2 and 3 show three bumps, indicated by arrows, at the positions where the areas in Figure \ref{inbetweenbox}(b1) were scanned. The accumulation of mass in the regions between the scanned areas was also found for higher doses, applied in two acquisitions and scanning with a dwell time higher than 9~$\mu\text{s}$; more accumulation was found when increasing the dose. 

\section{Discussion}
The above experiments can be summarised in the following statements which will serve as input for the second part of this manuscript which will attempt to qualitatively describe all observations from the simplest possible empirical model:
\begin{itemize}
\item{Damage occurs in regions that were not visited by the probe. This agrees with notions put forward by Egerton \cite{Egerton2012,EGERTON2017115} where the source of delocalised damage was interpreted as delocalised inelastic interaction with the sample.}
\item{Damage depends on the order in which the scan points are visited. This observation can not be explained by delocalised interaction and requires some notion of time dependent spreading of damage in space like e.g. locally applied heat would spread in a thin material.}
\item{Damage depends on the history of neighbouring areas. This requires some delocalised effect and the notion of damage building up with increasing dose.}
\item{Damage is higher in central regions of the image as these places have more neighbours. This requires a delocalised mechanism.}
\item{When doing mutiple aquisitions, the damage is lower if more frames are recorded for the same total time and same total dose. This links to dose fractionation experiments that were reported \cite{Jones2018} and again requires a notion of a damage mechanism that spreads in time. It has to be noted however that in our experiments the dose rate in each pixel was always kept constant so dose rate arguments alone can not explain this observation.}
\item{Damage seems to occasionally 'heal' with time. Occasionally we observed that damage could also be reverted in line with other observing similar effect \cite{Mkhoyan2006}. As an example, the self-filling of a hole after being created with the stationary beam is shown in Supporting Information video v1.}
\item{In Figure \ref{lineprofile}, similar line profiles were obtained only for the lowest dose applied during the first acquisitions, $3.17\times10^4~{e}^{-}\text{\AA}{}^{-2}$, with both scanning methods. For the further acquisitions in Figure \ref{lineprofile} or for the acquisitions in Figure \ref{lineprofile2}, with $4.76\times10^4~{e}^{-}\text{\AA}{}^{-2}$, differences in the line profiles are evident. This could be explained by a threshold effect. Although the same dose per frame is applied with both methods, the dose applied to a region of neighbouring pixels during time is different, in a 3 x 3 pixel region for instance. A more elaborated treshold effect can arise from the accumulated energy, even in one single pixel. As described before, the contribution to the energy in a region of the sample would not only have its origin on the actual scanned position but also on the neigbouring scanned regions. The alternative method may allow part of this energy to dissipate and a threshold value could be reached at a later step than with the raster method. This will be investigated in more detail in part II of this series.}
\end{itemize}
All of these observations hold clues for a more realistic damage model and the question arises whether other scan patterns or different scan parameters would allow for even lower beam damage while keeping the total dose constant. In order to answer this question, a numerical model is needed that mimicks all of these features and would allow searching for an optimal experimental design.
One could argue that the above observations are only relevant for our specific sample and indeed many if not all parameters in the model will be material dependent. However, our experiments show a systematic study with a statistically significant conclusion which leads to the identification of the important ingredients for a model that might well be applicable outside the current material class. After all the model will build on the dissipation of deposited energy without going into the details on how this process takes place in a specific material. We therefore expect that our observations can at least serve as a guide for the interpretation of beam damage in different materials.\\
In any case, our findings bring hope that there are ways to mitigate beam damage in STEM by adjusting the way the electron dose is applied to the sample. This might also contain a strong hint on why TEM is considered less prone to beam damage as compared to STEM, as indeed TEM could be seen as a fundamentally random sampled scanning as the electrons in the broad beam stochastically hit the sample in random positions (at least for inelastic interaction which is required for beam damage). This would give hope, especially for life science imaging, that the high contrast benefit of STEM (possibly with ptychography) can be combined with low beam damage, outperforming TEM based imaging modes. 

\section{Conclusion}
In this paper we have demonstrated that beam damage in a prototype commercial zeolite sample shows an interesting dependence on the applied scan strategy. At the same total dose and the same sampling of the images, a significant reduction of beam damage is observed for an alternative interleaved scanning pattern as compared to the traditional raster scanning pattern. We reach this conclusion through the application of a programmable scan engine that allows us to repeat multiple experiments under well-controlled conditions, providing a statistically relevant observation. These observations will serve as a basis for a second part of this manuscript that will attempt to build an empirical model that contains all ingredients to simulate all aspects of the experiments presented here. Such model could then allow us to make predictions on how to further  lower beam damage without necessarily lowering the electron dose.
Our observations support the (often vague) notion in the community that electron dose and acceleration voltage are not the only parameters affecting beam damage in (S)TEM experiments and rethinking the pattern in which this dose is applied holds promise to further shift the possibilities of EM for beam sensitive samples.

\section*{Acknowledgements}
A.V., D.J.,  A.B. and J.V. acknowledge funding from FWO project G093417N ('Compressed sensing enabling low dose imaging in transmission electron microscopy') and G042920N ('Coincident event detection for advanced spectroscopy in transmission electron microscopy'). This project has received funding from the European Union's Horizon 2020 research and innovation programme under grant agreement No 823717  ESTEEM3. The Qu-Ant-EM microscope was partly funded by the Hercules fund from the Flemish Government. J.V. acknowledges funding from GOA project “Solarpaint” of the University of Antwerp.

All data discussed in this manuscript is openly available through Zenodo to stimulate further research on this topic and to improve reproducibility.

%\section*{References}

\bibliography{P1_AV}





\end{document}

\section*{Supporting information}
%add this for the numbering of the figures in supplementary information
\beginsupplement

\begin{figure}[!h]
\includegraphics[width=\columnwidth]{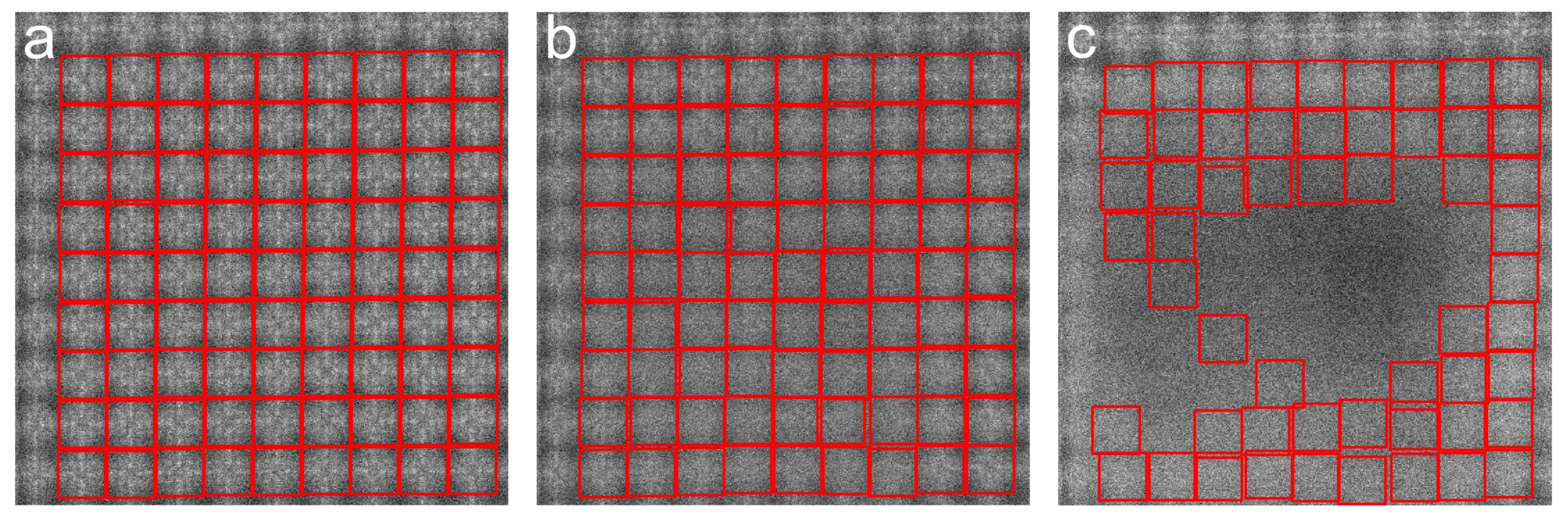}
\caption{\label{supplocationcell}Location of the cells (red squares) obtained with the cross-correlation method. The locations are shown on the raw raster images from Figure 4. (a), (b) and (c) correspond to the first, second and third acquisitions, respectively.}
\end{figure}

\begin{figure}[!h]
\includegraphics[width=\columnwidth]{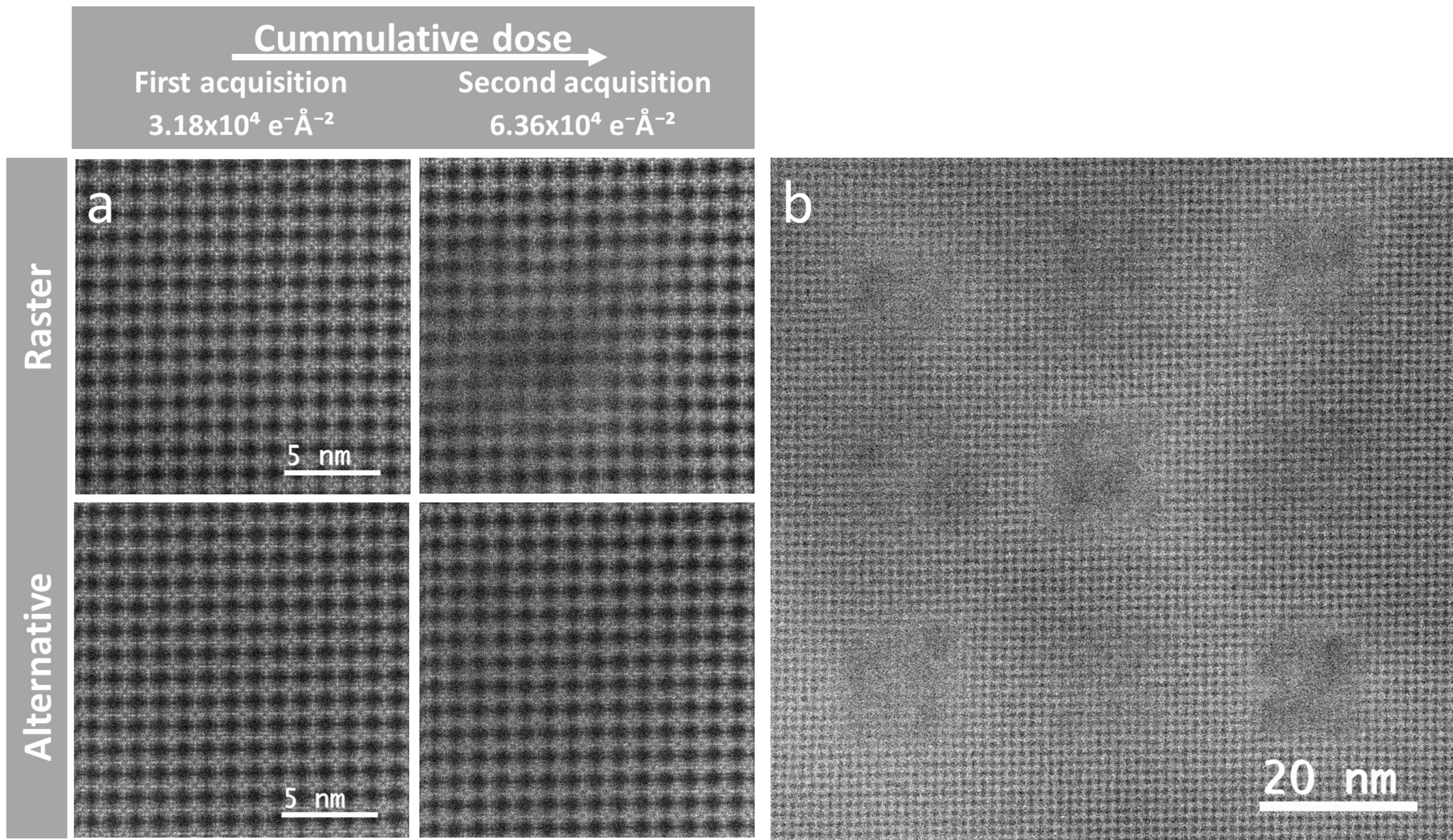}
\caption{\label{supphighrescomparison}Images from a $3\times3$ sub-images experiment consisting of two acquisitions performed at 12~$\mu\text{s}$ dwell time,  34.3~pm pixel size, and a calculated dose of approximately $3.18\times10^4~\text{e}^-\text\AA{}^{-2}$ per image. (a) Extracted images from the areas 5 and 4 (corresponding to the raster and alternative scanning, respectively) of the sub-images experiment, as depicted in Figure 2. Drift of the sample is observed because of the large dwell time. (b) Large field of view acquired after performing the experiment, image acquired at 3~$\mu\text{s}$ dwell time. The areas scanned with the alternative method show less damage than the areas scanned with the raster method.}
\end{figure}

\begin{figure}[!h]
\includegraphics[width=\columnwidth]{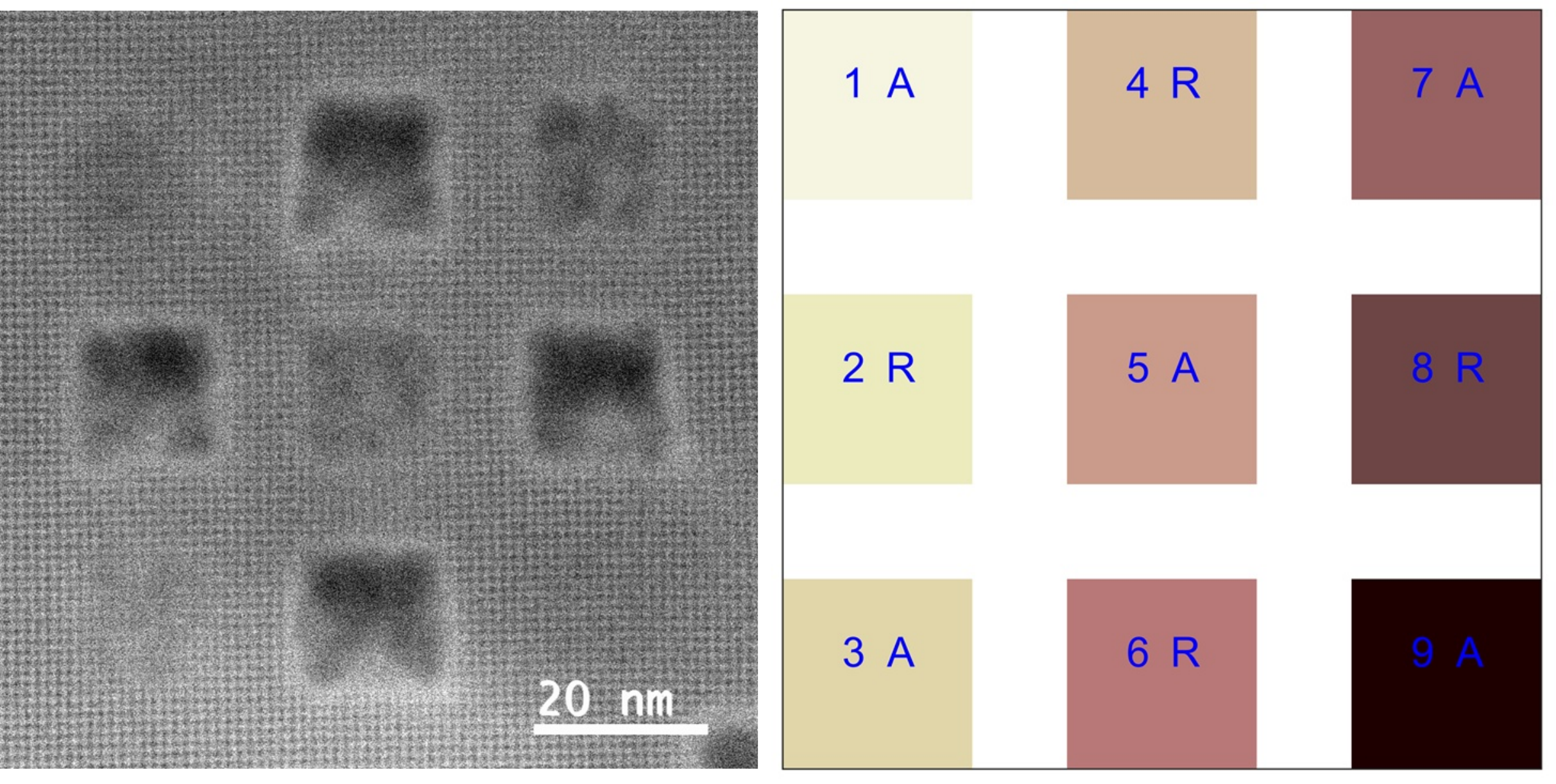}
\caption{\label{supplargefieldl}Left, large field of view acquired after performing an experiment consisting of two acquisitions acquired with 34.3~pm pixel size, 14~$\mu\text{s}$ dwell time and a calculated dose of approximately $3.71\times10^4~\text{e}^-\text\AA{}^{-2}$ per image. The sequence of the scanning methods for each squared area was inverted with respect to the previous results shown in the main manuscript. Right, sequence of the scanning methods employed here, alternating the alternative scanning (A) and raster scanning (R). The sequence is inverted with respect to the sequence on Figure 2. The areas scanned with the alternative method show less damage than the areas scanned with the raster method.}
\end{figure}

\begin{figure}[!h]
\includegraphics[width=\columnwidth]{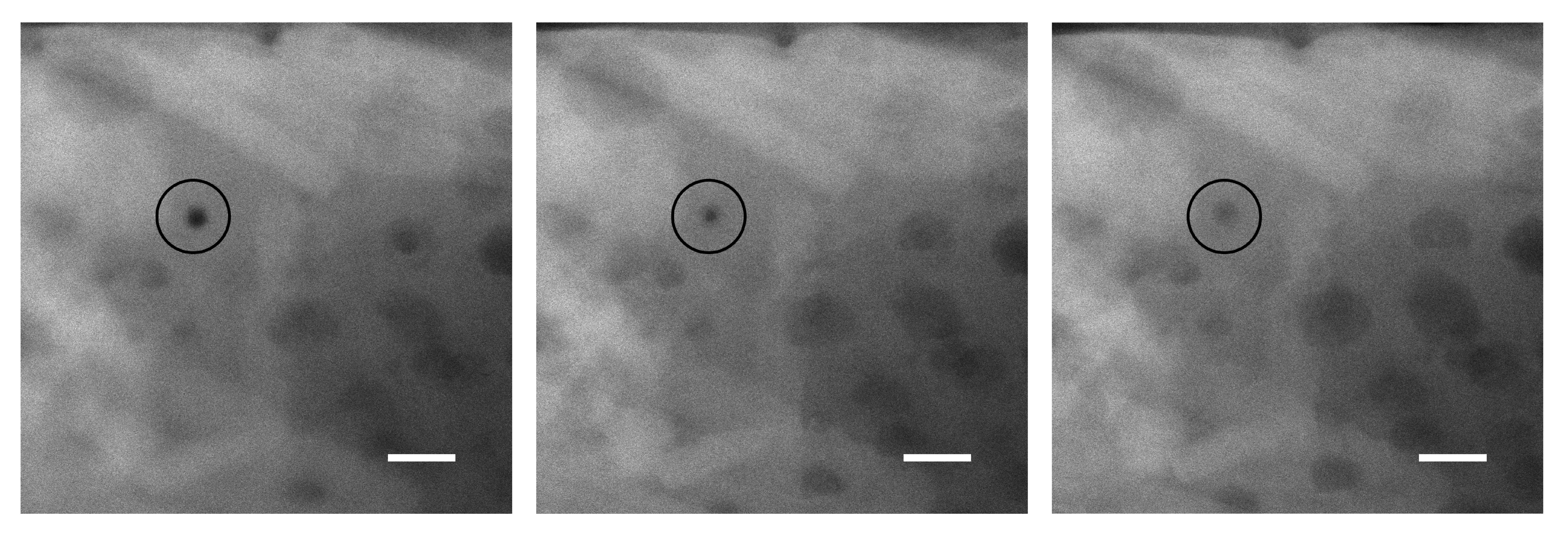}
\caption{\label{suppfilledhole}Partial self-filling of a hole after being created with a stationary beam. The hole is marked by the black circle. From left to right, the images were scanned with a time interval of approximately 38 s between the acquisitions. The scale bar represents 20~nm. See video v1 in Supporting Information.}
\end{figure}

%\section{Experimental}

%\section{Results}

%\section{Discussion}

%\section{Conclusion}

%\section*{Acknowledgements}

%\section*{References}

%\bibliography{P1_AV}



